\pdfoutput = 1
\documentclass[apj,numberedappendix]{emulateapj}

\usepackage{epsfig}
\usepackage{amsmath}
\usepackage{enumerate}
\usepackage{mathrsfs}
\usepackage{natbib}
\usepackage{units}
\usepackage{hyperref}

\usepackage{ulem}
\usepackage[squaren, Gray, cdot]{SIunits}

\usepackage{color}
\definecolor{orange}{RGB}{220,110,0}

\newbox\grsign \setbox\grsign=\hbox{$>$} \newdimen\grdimen \grdimen=\ht\grsign
\newbox\simlessbox \newbox\simgreatbox \newbox\simpropbox
\setbox\simgreatbox=\hbox{\raise.5ex\hbox{$>$}\llap
     {\lower.5ex\hbox{$\sim$}}}\ht1=\grdimen\dp1=0pt
\setbox\simlessbox=\hbox{\raise.5ex\hbox{$<$}\llap
     {\lower.5ex\hbox{$\sim$}}}\ht2=\grdimen\dp2=0pt
\setbox\simpropbox=\hbox{\raise.5ex\hbox{$\propto$}\llap
     {\lower.5ex\hbox{$\sim$}}}\ht2=\grdimen\dp2=0pt
\def\simgt{\mathrel{\copy\simgreatbox}}
\def\simlt{\mathrel{\copy\simlessbox}}

\newcommand{\md}{\mathrm{d}}
\newcommand{\alfven}{Alfv\'{e}n }

\newcommand{\bb}{\boldsymbol{B}}
\newcommand{\be}{\boldsymbol{E}}
\newcommand{\bj}{\boldsymbol{J}}
\newcommand{\bp}{\boldsymbol{P}}
\newcommand{\bF}{\boldsymbol{F}}
\newcommand{\bA}{\boldsymbol{A}}
\newcommand{\bz}{\boldsymbol{z}}
\newcommand{\bV}{\boldsymbol{v}}
\newcommand{\fdiss}{f_{\rm diss}}
\newcommand{\fesc}{f_{\rm esc}}

\begin{document}

\title{Dissipation of Alfv\'{e}n Waves in Relativistic Magnetospheres of Magnetars}

\author{Xinyu Li\altaffilmark{1}}
\author{Jonathan Zrake\altaffilmark{1}}
\author{Andrei M. Beloborodov\altaffilmark{1,2}}

\affil{$^1$Physics Department and Columbia Astrophysics Laboratory, Columbia University, 
538 West 120th Street, New York, NY 10027\\
$^2$Max Planck Institute for Astrophysics, Karl-Schwarzschild-Str. 1, D-85741, Garching, Germany}

\begin{abstract}
Magnetar flares excite strong \alfven waves in the magnetosphere of the neutron star. 
The wave energy can (1) dissipate in the magnetosphere, (2) convert to ``fast modes'' and possibly escape, and (3) penetrate the neutron star crust and dissipate there.
We examine and compare the three options. Particularly challenging are nonlinear interactions between strong waves, which develop a cascade to small dissipative scales. 
This process can be studied in the framework of force-free electrodynamics (FFE).
We perform three-dimensional FFE simulations to investigate \alfven wave dissipation, how long it takes, and how it depends on the initial wave amplitude on the driving scale. In the simulations, we launch two large \alfven wave packets that keep bouncing on closed magnetic field lines and collide repeatedly until the full turbulence spectrum develops.
Besides dissipation due to the turbulent cascade, we find that in some simulations spurious energy losses occur immediately in the first collisions. 
This effect occurs in special cases where the FFE description breaks. It is explained with a simple one-dimensional model, which we examine in both FFE and full magnetohydrodynamic settings.
We find that magnetospheric dissipation through nonlinear wave interactions is relatively slow and more energy is drained into the neutron star. The wave energy deposited into the star is promptly dissipated through plastic crustal flows induced at the bottom of the liquid ocean, and a fraction of the generated heat is radiated from the stellar surface.
\medskip

\end{abstract}

\keywords{dense matter --- magnetic fields --- stars: magnetars --- stars: neutron --- waves}

\section{Introduction}
Magnetars are neutron stars hosting ultra-strong magnetic fields of $10^{14}-10^{16}$~G (see \citet{2017ARA&A..55..261K} for a recent review).
They exhibit a broad range of X-ray activity, including giant flares with luminosities up to $10^{47}$~erg/s.
The flares are likely powered by a sudden magnetospheric rearrangement that dissipates magnetic energy \citep{1995MNRAS.275..255T,1996ApJ...473..322T}.
A slower mode of dissipation is invoked to explain persistent hard X-ray emission \citep{2013ApJ...762...13B}.
Magnetic energy dissipation in the magnetosphere generally plays a key role in magnetar activity.

One proposed dissipation mechanism is the turbulent cascade of magnetospheric \alfven waves excited by a starquake \citep{1996ApJ...473..322T}. \alfven waves are also excited when the magnetosphere is slowly ``overtwisted'' and loses equilibrium, as observed in simulations by \citep{2013ApJ...774...92P}. The excited waves can have large amplitudes and carry a significant fraction of the magnetospheric energy. 

\citet{1996ApJ...473..322T} proposed that the waves will cascade to small dissipative scales and convert to heat, creating an energetic ``fireball'' of thermalized $e^\pm$ plasma. However, there are two competing processes that can remove the wave energy. First, \alfven waves can convert to so-called ``fast modes'' capable of escaping the magnetosphere. Unlike \alfven waves, which are ducted along the magnetic field lines and trapped in the closed magnetosphere, fast modes can propagate across the field lines.
Secondly, \citet{2015ApJ...815...25L} showed that \alfven waves bouncing in the magnetosphere are gradually drained into the stellar crust, where they initiate plastic flows and dissipate. About 10 bouncing cycles are typically sufficient to damp the waves by this mechanism, and \citet{2015ApJ...815...25L} suggested that this may occur faster than dissipation of waves through a turbulent cascade in the magnetosphere. Evaluating the efficiency of the latter process requires a detailed calculation of nonlinear processes, which can be done numerically. We attempt this calculation in the present paper.

The theory of turbulent cascades has a long history. The MHD cascade is different from the hydrodynamic cascade where energy transfer is mediated by interacting vortices. The difference is seen already in the simplest, incompressible, non-relativistic MHD, where only \alfven waves are present. In the case of weak turbulence (meaning that the time for energy transfer across spatial scales is longer than the wave period),
the three-wave interaction is prohibited by kinetic constraints \citep{1994ApJ...432..612S}. 
Then nonlinear interactions are dominated by the four-wave interactions among \alfven waves and give rise to the anisotropic energy cascade in the direction perpendicular to the background field. For strong turbulence, a $k_\perp^{-5/3}$ spectrum was predicted from detailed balance \citep{1995ApJ...438..763G}.

More work was done later to include compressive modes -- the fast and slow magnetosonic waves.
Using the random phase approximation \citet{2001JETP...93.1052K} found a weak-turbulence spectrum $k_{\perp}^{-2}$.
Relativistic nonlinear Alfv\'enic turbulence was studied using numerical simulations \citep{2005ApJ...621..324C,2012ApJ...744...32Z,2016ApJ...817...89Z,2016ApJ...831L..11T,2017MNRAS.472.4542T}, but far less than in the non-relativistic setting. 
Simulations by \citet{2016ApJ...831L..11T,2017MNRAS.472.4542T} suggested that compressible modes are strongly coupled with \alfven waves and participate in the energy cascade.

Similar to the previous works, we are interested in low-frequency waves, which are described by relativistic magneto-hydrodynamics (RMHD). In the magnetically-dominated limit (negligible plasma inertia), a simpler approximation of ``force-free electrodynamics'' (FFE) becomes useful. In this approximation, the plasma energy and momentum are neglected and the stress-energy tensor of the electromagnetic field $T^{\mu\nu}$ satisfies $\nabla_\mu T^{\mu\nu}=0$.
Both RMHD and FFE support \alfven waves, which transport energy along the direction of the background field, and also support the fast modes.\footnote{In FFE, name ``fast'' is somewhat misleading, because all waves propagate with the speed of light.}

An analytical study of wave interaction in FFE by \citet{1998PhRvD..57.3219T} found that an \alfven wave pair can convert to a fast wave via three-wave interactions. 
In contrast to \alfven waves, the group velocity of the fast modes can be in any direction and they can possibly escape the magnetosphere, carrying energy away.

In this paper we use FFE and RMHD simulations to investigate the efficiency of nonlinear processes in removing wave energy through dissipation and escape.
Our goals are to determine the fate of wave energy in the context of giant magnetar flares, and to study the nonlinear dynamics of interacting \alfven waves from a physics perspective.
We present numerical simulations of relativistic \alfven wave turbulence operating in a toy magnetosphere replaced by a rectangular box.
We utilize a high-order conservative finite differencing scheme to evolve the FFE equations, and devote particular attention to the code's modeling of energy dissipation. We discuss the various modes by which energy is removed numerically, and point out (via direct comparison with a relativistic MHD code) circumstances when FFE wrongly models the energy dissipation rate.

Our paper is organized as follows. In Section 2, we give a brief description of wave modes and nonlinear interactions in FFE. In Section 3, we outline our numerical scheme and discuss energy dissipation channels it admits. 2D and 3D numerical results for \alfven wave turbulence driven by colliding wave packets are presented in Section 4. 
In Section 5, we show that FFE simulations can badly over-predict the energy dissipation rate, as the result of a commonly employed technique for maintaining magnetic dominance ($E < B$). The final section is devoted to a discussion of our results in the context of the fireball model for magnetar giant flares. Throughout this paper we utilize units in which speeds are measured in units of the speed of light $c$, and electric ($\boldsymbol{E}$) and magnetic ($\boldsymbol{B}$) field values are normalized by $\sqrt{4 \pi}$.

\section{Waves and Nonlinear Interactions in FFE}
\subsection{Equations of FFE}

FFE describes relativistic magnetically dominated plasma, where the plasma inertia can be neglected, i.e. $\rho \ll B^2/2$ where $\rho$ is the mass density of plasma. The dynamical equations are given by Maxwell's equations,
\begin{eqnarray}\label{Maxwell}
	\frac{\partial\boldsymbol{B}}{\partial t}+\nabla\times \boldsymbol{E},
    \qquad
	\frac{\partial\boldsymbol{E}}{\partial t}-\nabla\times \boldsymbol{B} &=& -\boldsymbol{J} \, ,
\end{eqnarray}
together with 
the vanishing force condition $\nabla_\mu T^{\mu\nu}=0$ or
\begin{equation}\label{ffe_condition}
	\rho_e \boldsymbol{E}+\boldsymbol{J}\times \boldsymbol{B} = 0 \, ,
\end{equation}
where $\rho_e=\nabla\cdot\be$ is the charge density. The force-free condition, Equation \ref{ffe_condition}, requires $\be \cdot \bb = 0$ and $E < B$.
Equation \ref{ffe_condition} and $\partial_t(\be \cdot \bb) = 0$ together yield the the following expression for the electric current density (e.g. \citet{2002MNRAS.336..759K}),
\begin{equation}\label{current}
	\bj=\boldsymbol{J}_{\rm FFE} \equiv \rho_e \frac{\boldsymbol{E}\times\boldsymbol{B}}{B^2}+\frac{\boldsymbol{B}\cdot\nabla\times\boldsymbol{B}-\boldsymbol{E}\cdot\nabla\times\boldsymbol{E}}{B^2}\boldsymbol{B} \, .
\end{equation}
$\bj_{\rm FFE}$ introduces nonlinearity into the Maxwell equations.

Since FFE neglects the plasma energy, the total energy of the system is given by
\begin{equation}
	U_{\rm tot} = \int\md V\;\frac{1}{2}(B^2+E^2) \, .
\end{equation}
This energy is formally conserved because Equation \ref{ffe_condition} guarantees $\be \cdot \bj = 0$.

\subsection{Wave solutions in FFE}
\label{waves}
We will use the temporal gauge where the electric scalar potential $\varphi$ is set to zero,
and the vector potential $\bA$ fully specifies the electromagnetic field,
\begin{eqnarray}\label{potential}
	\boldsymbol{B} = \nabla\times \bA, \qquad \boldsymbol{E} = - \frac{\partial \bA}{\partial t} \, .
\end{eqnarray}
The Maxwell equations then reduce to 
\begin{equation}\label{reducedMax}
	\frac{\partial^2\bA}{\partial t^2} + \nabla\times \nabla\times\bA = \bj \, .
\end{equation}

We approximate the steady background magnetic field $\bb^{(0)}$ as uniform (i.e. limit our consideration to waves much shorter than the variation scale of the background field), and choose the $z$-axis along $\bb^{(0)}$ and the $y$-axis along $\bA^{(0)}$,
\begin{equation}
  \bA^{(0)}=B_{0}x\,\hat{\boldsymbol{y}}, \qquad 
 \bb^{(0)} = B_{0}\, \hat{\boldsymbol{z}},   \qquad \boldsymbol{E}^{(0)}=0.
\end{equation} 
$A^{(0)}$ has no time dependence, and so there is no background electric field.

Approximate solutions for waves and their interactions may be obtained by use of a perturbative expansion,
\begin{eqnarray}\label{pert}
	\bA = \bA^{(0)}
    +\epsilon \bA^{(1)}+\epsilon^2\bA^{(2)}+\cdots,
\end{eqnarray}
where $\epsilon\ll 1$.
We seek solutions for the perturbed quantities of the form 
\begin{equation}
\label{Fourier}
\bA^{(n)}(t,\boldsymbol{r}) \propto \exp[i(\boldsymbol{k}^{(n)} \cdot \boldsymbol{r} - \omega^{(n)} t)]\, ,
 \qquad n\geq 1,
\end{equation}
where $\boldsymbol{r}=(x,y,z)$ is the position vector.

Inserting Equation~(\ref{potential}) into the expression for $\bj$ (Equation~\ref{current}), substituting the result into Equation~\ref{reducedMax}, and keeping only terms up to the first order in $\epsilon$ yields the  linear equation for $\bA^{(1)}$ of the form
\begin{equation}\label{Maxwell1}
   \mathcal{L}[\bA^{(1)}]=0,
\end{equation} 
where 
\begin{equation}
  \mathcal{L}\equiv \frac{\partial^2}{\partial t^2} + (\nabla\times \nabla\times)_\perp
\end{equation}
is a linear differential operator. The operator becomes algebraical when it is applied to the Fourier modes (Equation~\ref{Fourier}), and the wave equation becomes $L\bA^{(1)}=0$, where $L(\omega,\boldsymbol{k})$ is a matrix. The condition $\det L=0$ for the existence of solutions $\bA^{(1)}\neq 0$ gives two pairs of roots $\omega(\boldsymbol{k})$, which describe the dispersion relations of the propagating eigen modes. The corresponding eigen vectors $\boldsymbol{e}_m$ represent the wave polarization, and each eigen mode may be written in the form
\begin{equation}
  \bA^{(1)}_m=\Lambda_m\,\boldsymbol{e}_m,
\end{equation}
where $\Lambda_m$ represents the wave amplitude.

For any Fourier mode the induction equation $\partial \bb/\partial t=-\nabla\times\be$ implies $\omega\bb=\boldsymbol{k}\times\be$, and hence the condition $\be\cdot\bb=0$ is automatically satisfied. In our setting, the first-order expansion of $\be\cdot\bb =\be^{(0)}\cdot\bb^{(1)}+\be^{(1)}\cdot\bb^{(0)}\propto \bA^{(1)}\cdot\bb^{(0)}$ implies
\begin{equation}
\label{eq:Az}
  A^{(1)}_z=0,
\end{equation}
i.e. the polarization vectors $\boldsymbol{e}_m$ must be perpendicular to the background magnetic field. 

A straightforward calculation shows that two distinct modes are supported by FFE:
\begin{enumerate}
	\item \alfven wave ---
	This mode has the dispersion relation $\omega(\boldsymbol{k})=\pm k_z$ and the polarization vector
	\begin{equation} \label{polA}
		\boldsymbol{e}_\mathcal{A} = \frac{\boldsymbol{k}_\perp}{\sqrt{\omega}|\boldsymbol{k}_\perp|}\,,
	\end{equation}
	where $\boldsymbol{k}_\perp$ is the component of wave vector perpendicular to the background field $\bb^{(0)}$.
	The electric field in the wave $\be^{(1)}=-i\omega\bA^{(1)}$ is along $\boldsymbol{k}_\perp$, and the magnetic field $\bb^{(1)}=i\boldsymbol{k}\times\bA^{(1)}$ is along $\hat{\boldsymbol{z}}\times\boldsymbol{k}_\perp$.
	\alfven waves have group velocity along $\pm \hat{\boldsymbol{z}}$, and therefore can only transport energy parallel (or anti-parallel) to the background field. The sign in the dispersion relation indicates the direction of the wave. The current associated with \alfven waves is
    $\bj_{\mathcal{A}} \propto
    k_\perp \sqrt{\omega} \hat{\boldsymbol{z}}$, which is non-zero for $k_\perp \ne 0$.

	\item Fast wave ---
	The dispersion relation is $\omega(\boldsymbol{k})=\pm |\boldsymbol{k}|$ with the polarization vector
    \begin{equation} \label{polF}
	\boldsymbol{e}_\mathcal{F} = \frac{\boldsymbol{k}_\perp\times\hat{\boldsymbol{z}}}{\sqrt{\omega}|\boldsymbol{k}_\perp|} \, .
	\end{equation}
    Then $\be^{(1)}$ is along $\boldsymbol{k}_\perp\times\hat{\boldsymbol{z}}$, and $\bb^{(1)}=(\boldsymbol{k}\times\be^{(1)})/\omega$ is in the $\boldsymbol{k}_\perp$-$\hat{\boldsymbol{z}}$ plane and perpendicular to $\boldsymbol{k}$.
	Fast waves in FFE create no charge density $\rho_e=\nabla\cdot \be^{(1)}=i\boldsymbol{k}\cdot \be^{(1)}=0$, and also no current density, $\bj_{\mathcal{F}}=0$.
    Therefore, the fast waves propagate as vacuum electromagnetic waves.
\end{enumerate}

When $k_\perp=0$, the two wave modes become degenerate. Notably, while these wave solutions have been derived from the linearized equations, they are in fact exact nonlinear solutions to the FFE equations.

The polarization vectors $\boldsymbol{e}_{\mathcal{A},\mathcal{F}}$ in Equations \ref{polA} and \ref{polF} are normalized so that the energy of an ensemble of fast and \alfven waves takes the form
\begin{equation}
	U =\sum\limits_{m=\mathcal{A},\mathcal{F}} \sum\limits_{\boldsymbol{k}}\omega_m
\Lambda_m^\star(\boldsymbol{k})
\Lambda_m(\boldsymbol{k}).
\end{equation}

\subsection{Wave-wave interactions}
Nonlinear interactions between waves arise from the current density $\bj$.
The lowest order interaction involves three waves, where two waves generate a third. These three-wave interactions are identified by inserting the expansion for two modes, $\bA^{(1)} = \bA^{(1)}_1 + \bA^{(1)}_2$, into Maxwell's equations, and equating the second order terms,
\begin{eqnarray}\label{Maxwell2}
	\mathcal{L}[\bA^{(2)}] =  \bj^{(2)}_{\rm nl} \, ,
\end{eqnarray}
The second order term $\bj^{(2)}_{\rm nl}$ in the-force free current is cumbersome and presented in Appendix. It is instructive to consider the following variants of the incoming waves $\bA^{(1)}_1+\bA^{(1)}_2$.

For two incoming fast modes one finds that $\boldsymbol{J}_{\rm nl}^{(2)}\neq 0$ is possible (in contrast to their $\boldsymbol{J}^{(1)}=0$). However in this case, $\boldsymbol{J}_{\rm nl}^{(2)}$ is parallel to the guide field $\bb^{(0)}$, and  sources $\bA^{(2)}$ along $\hat {\boldsymbol{z}}$. 
There are no propagating modes with $A_z\neq 0$ (see Equation~(\ref{eq:Az})), and so the three-wave interaction with two incoming fast modes is suppressed. 

For two incoming \alfven waves propagating in the same direction along the guide field ($k^{(1)}_{1,z}$ has the same sign as $k^{(1)}_{2,z}$), one finds that $\bj^{(2)}_{\rm nl}$ vanishes. Therefore, only counter-propagating \alfven waves can generate new waves through 3-wave interaction.
The generated wave has wavevector $\boldsymbol{k}^{(2)} = \boldsymbol{k}^{(1)}_1+ \boldsymbol{k}^{(1)}_2$ and frequency $\omega^{(2)} = \omega^{(1)}_1+\omega^{(1)}_2$. 
The excitation of the second-order wave is enhanced for the resonant three-wave interaction, meaning that $\bA^{(2)}$ is also a linear eigen mode. One can show that $\boldsymbol{k}^{(2)}$ and $\omega^{(2)}$ may satisfy the dispersion relation of \alfven waves only if
one of the incoming waves has $k_z = 0$, and such modes do not propagate, as they have $\omega=0$ according to the dispersion relation $\omega=\pm k_z$.
Therefore, two counter-propagating \alfven waves can only participate in resonant interactions where the outgoing wave is a fast mode ($\mathcal{A}+\mathcal{A}\rightarrow \mathcal{F}$).

Resonant three-wave interactions are also possible between an incoming \alfven wave and an incoming fast wave, and the outgoing wave can be either a fast wave or an \alfven wave ($\mathcal{A}+\mathcal{F}\rightarrow \mathcal{A/F}$).

\section{Numerical setup}
\subsection{Computational setting}
We perform numerical simulations in a fully periodic domain with uniform guide magnetic field aligned with the $z$-axis, $B_z=1$.
The box extends from 0 to 1 along each axis, and the wave crossing time is also unity.
We utilize initial conditions comprising a pair of counter-propagating \alfven wave packets.
Due to the use of periodic boundary conditions, the wave packets collide repeatedly, twice each time they traverse the computational domain; the time interval between successive collisions is $\tau = 0.5$.
This setting simulates \alfven wave packets propagating on closed field lines anchored on the magnetar surface, neglecting geometric effects due to the field-line curvature. The periodic boundary condition corresponds to an idealized situation where \alfven waves are perfectly reflected when hitting the magnetar surface.

\subsection{Solution scheme}
\label{scheme}
We numerically evolve the FFE equations using a third-order in time, fifth-order in space, flux-conservative scheme based on the WENO method \citep{doi:10.1137/070679065} adapted to FFE \citep{2011MNRAS.411.2461Y}.
We define a vector of primitive variables,
\begin{equation}
	\bp= \left(B_x,B_y,B_z, E_x,E_y,E_z\right)^\top \, ,
\end{equation}
and rewrite the Maxwell equations in the flux-conservative form
\begin{equation} \label{maxwell-fv}
	\partial_t \bp + \partial_x\bF^x + \partial_y\bF^y + \partial_z\bF^z = \boldsymbol{T} \, ,
\end{equation}
where the source term $\boldsymbol{T}$, and the flux functions are given by
\begin{eqnarray}
	\boldsymbol{T} &=& \left(0,0,0, -J_x, -J_y, -J_z\right)^\top \nonumber\\
	\bF^x &=& \left(0,-E_z,E_y,0,B_z,-B_y\right)^\top \nonumber\\
	\bF^y &=& \left(E_z,0,-E_x,-B_z,0,B_x\right)^\top \nonumber\\
	\bF^z &=& \left(-E_y,E_x,0, B_y,-B_x,0 \right)^\top \, .
\end{eqnarray}
The components of electric current appearing in $\boldsymbol{T}$ are computed using standard forth order finite differencing on the volume-centered values of $\be$ and $\bb$, according to Equation \ref{current}.

Time stepping is accomplished using the third-order TVD Runge-Kutta (RK) method \citep{Gottlieb:1998:TVD:279724.279737}.
Each RK sub-step updates the primitive variables by adding the source term and face-centered fluxes in a finite-volume form of Equation \ref{maxwell-fv}
\begin{equation}
    \bp^{n + 1} = \bp^n + \boldsymbol{T}^n \Delta t - \frac{\Delta t}{\Delta V} \sum_{\rm{faces}} \Delta S_i \hat{\boldsymbol{F}}^i\, .
\end{equation}
Here $\hat{\boldsymbol{F}}^i$ are the face-centered fluxes, evaluated from $\bp^{n}$, and Roe's Riemann solver,
\begin{eqnarray}
	\hat{\bF}^i_{j+1/2} &=& \frac{1}{2}\Big[\bF^i(\bp^+_{j+1/2}) + \bF^i(\bp^-_{j+1/2}) \nonumber\\
    && -\sum\limits_{m=1}^6  \left|\lambda^i_{m, j+1/2}\right| \alpha^i_{m, j+1/2}\boldsymbol{v}^i_{m, j+1/2} \Big]
\end{eqnarray}
where  $\lambda^i_{m, j+1/2}$ and $\boldsymbol{v}^i_{m, j+1/2}$ are eigenvalues and eigenvectors of the Jacobian matrix $\partial \bF^i / \partial \bp$,
and
\begin{equation}
	\alpha^i_{m, j+1/2,j,k} = \left(\bp^+_{j+1/2}-\bp^-_{j+1/2}\right) \cdot \boldsymbol{v}^i_{m, j+1/2}
\end{equation}
is the projection of the difference between left and right states to the eigenvectors.
The left and right states $\bp^{\pm}$ are reconstructed from $\boldsymbol{P}^n$ using the fifth order WENO method. 

\subsection{Constraint preservation}
\label{constraint}
In order to keep the magnetic field divergence-free, we utilize the hyperbolic divergence cleaning approach outlined in \citet{2002JCoPh.175..645D}. In practice, we find this approach maintains $\nabla \cdot \bb = 0$ to high precision. Note that violations in $\nabla \cdot \bb = 0$ only arise from the truncation error of the numerical scheme, and so the modifications to the solution introduced by the hyperbolic cleaning step converge away with increasing numerical resolution. Hyperbolic divergence cleaning involves the addition of an auxiliary scalar field $\Psi$ and a corresponding evolution equation. For brevity, we have excluded this equation from the description of our numerical scheme in Section \ref{scheme}. For details we refer the reader to \cite{2002JCoPh.175..645D}.

Small violations in the $\be \cdot\bb = 0$ constraint also arise at the level of truncation error. Instead of removing the parallel component of $\be$ at each time step, we introduce a correction term to the force-free Ohm's law that allows for time-resolved damping of $\be_\parallel$ \citep{2017MNRAS.469.3656P},
\begin{eqnarray} \label{modified-ffe-ohm}
		\bj_m &=& \rho_e \frac{\be\times\bb}{B^2 +\tilde{E}^2} \nonumber\\
		 && + \frac{\bb\cdot\nabla\times \bb - \be\cdot\nabla\times\be + \gamma \be\cdot\bb}{B^2}\bb \, .
\end{eqnarray}
Here, $1 / \gamma$ is a time scale for the damping of $\be_\parallel$ (typically chosen to be several times $\Delta t$), and the modified electric field magnitude $\tilde E^2$ appearing in the denominator of the first term in Equation \ref{modified-ffe-ohm} is defined as \citep{2012ApJ...746...60L}
\begin{equation}
	\tilde{E}^2 = \frac{1}{2} \left(\sqrt{\chi^4 + 4 \be \cdot \bb} + \chi^2\right) - \chi^2 \, ,
\end{equation}
where $\chi^2 \equiv B^2 - E^2$.
When $\be \cdot \bb = 0$, the modified current density Equation \ref{modified-ffe-ohm} reduces to Equation \ref{current}.

\subsection{Maintaining magnetic dominance}
\label{magnetic-dominance}
Self-consistent evolution of the FFE equations requires magnetic dominance ($E < B$) to be maintained. However, non-linear FFE solutions in which $E$ remains everywhere smaller than $B$ generally exist only for finite time. This reflects that realistic plasma systems, having small but finite rest-mass energy, inevitably develop regions where the thermal pressure gradient or the MHD inertial term becomes important.
Such conditions arise either where $B^2 / 2$ drops below $\rho$ (e.g. near magnetic null points), or where plasma is accelerated to high Lorentz factor. In such regions the electric current deviates significantly from $\bj_{\rm FFE}$, enabling momentum transfer between the plasma and the electromagnetic field. 

A standard procedure to continue numerical evolution of the FFE equations is to artificially reduce the magnitude of $E$ wherever $E > B$,
\begin{equation}\label{shrinkE}
	\be\rightarrow\sqrt{\frac{B^2}{E^2}} \be \, .
\end{equation}
This procedure is commonly interpreted as modeling a dissipative process \citep{2006MNRAS.367.1797M, 2006ApJ...648L..51S}, such as the rapid acceleration of charged particles enabled by $E > B$. However, violation of the force-free condition in real plasma systems does not necessarily lead to energy dissipation. We will demonstrate this explicitly in Section \ref{sec:breakffe} by comparing our FFE solutions with those of strongly magnetized relativistic MHD. The MHD solutions reveal that breaking of the force-free condition leads to time-reversible momentum exchange between the field and the plasma. We will thus conclude that electromagnetic dissipation is not properly modeled by FFE when significant energy is lost as a consequence of the procedure in Equation \ref{shrinkE}.

\subsection{Dissipation channels in FFE simulations}
\label{dissipation-channels}
Although FFE formally conserves energy, numerical evolution schemes require some dissipation in order to maintain stability, satisfy constraints, and keep the solution magnetically dominated. There are four channels for energy dissipation in our numerical simulations:
\begin{enumerate}[(i)]
	\item The hyperbolic divergence cleaning step, which leads to 
	\begin{equation}
		\partial_t U = -\bb \cdot \nabla \Psi \, ,
	\end{equation}
	where $\Psi$ is the auxiliary scalar function discussed in Section \ref{constraint}.
	\item Dissipation introduced by the modified force-free current $\bj_m$ in Equation \ref{modified-ffe-ohm}.
	\begin{eqnarray}
	 \partial_t U &=& - \bj_m \cdot\be \nonumber \\ 
	                   &=& -(\be\cdot\bb)\frac{\bb\cdot\nabla\times \bb-\be\cdot\nabla\times\be}{B^2} \nonumber \\
	                   &&-\frac{\gamma (\be\cdot\bb)^2}{B^2}.
	\end{eqnarray}
	\item Reduction of electrical field when $E>B$ (Equation \ref{shrinkE}).
	\item Subtraction of short-wavelength field oscillations at the grid scale, referred to as ``grid heating.''
\end{enumerate}

Channels (i) and (ii) become less significant with increasing grid resolution, because the numerical values of $\Psi$ and $\be\cdot\bb$ are proportional to the truncation error of the numerical scheme. In the results presented in Sections \ref{sec:spec} and \ref{sec:fate}, these channels do not contribute significantly to the measured energy dissipation rate.

Channel (iii) does not in general become small as the grid resolution increases. 
As mentioned in Section~\ref{magnetic-dominance}, this dissipation may be artificially strong.
Therefore, we consider our measurements of the energy dissipation rate to be reliable only when dissipation is not dominated by this effect.

Channel (iv), energy removal by grid heating, may or may not ``converge away'' with increasing resolution. For example, the FFE wave solutions discussed in Section \ref{waves} evolve without any significant grid heating, provided their wavelength is well resolved. Generally, the rate of grid heating of isolated waves will depend on the numerical resolution. In contrast, non-linear numerical solutions can exhibit significant energy loss over time, at a rate that becomes independent of grid resolution. Such behavior usually reflects the presence of a forward energy cascade, in which the rate of high frequency wave damping is determined by the rate of energy transfer into high frequency modes by numerically resolved non-linear interactions. In such cases, grid heating is expected to capture the true dissipation rate.

\subsection{Numerical diagnostics}
A useful diagnostic in our analysis will be the free energy $U$, which we define to be the total electromagnetic energy of the system, but with the contribution from the background magnetic field $B_z$ removed,
\begin{equation}
	U \equiv U_{\rm tot} - \frac{1}{2} \int \md V\; B_z^2 \, .
\end{equation}

We will also utilize the power spectra $P(k_\parallel)$ and $P(k_\perp)$, representing the distribution of electromagnetic energy in wavenumber parallel and perpendicular to the background field. The power spectra are obtained by binning the square of the discrete Fourier modes $\tilde{\bb}(\boldsymbol{k})$ and $\tilde{\be}(\boldsymbol{k})$ by the wavenumber components $k_\parallel$ and $k_\perp$,
\begin{equation} \label{eqn:power-spectrum}
	P(k_{\parallel,\perp}) \Delta k_{\parallel,\perp}\equiv \frac{1}{2 U_0} \sum\limits_{\boldsymbol k \in \Delta k_{\parallel, \perp}} |\tilde{\bb}(\boldsymbol{k})|^2 + |\tilde{\be}(\boldsymbol{k})|^2 \, .
\end{equation}
Note that the spectra are normalized to the free energy $U_0$ in the system at $t=0$, such that
\begin{equation}
	U = U_0 \sum 
	P(k_\parallel)\Delta k_\parallel =
    U_0 \sum 
      P(k_\perp) \Delta k_\perp \, .
\end{equation}
In 2D and 3D runs, the spectral bins $\Delta k_\parallel$ are planar slabs orthogonal to the background field. The spectral bins $\Delta k_\perp$ are planar slabs in 2D and cylindrical annuli in 3D.

\section{Spectral evolution of wave turbulence}\label{sec:spec}
\subsection{3D simulations}\label{sec:3d}
In order to study the interaction between \alfven modes in a three-dimensional setting, we initialize two counter-propagating wave packets, each perturbing the background field within a spherical volume. We utilize initial conditions
$\bb = B_0\hat{\boldsymbol{z}}+ \nabla \times(\phi \hat{\boldsymbol{z}})$ with scalar field
\begin{equation}\label{3dpacket}
	\phi = \xi\ell\sum\limits_{i=1,2} \exp\left(- \frac{|\boldsymbol{r}-\boldsymbol{r}_i|^2}{\ell^2} \right),
\end{equation}
where $\boldsymbol{r}_1 = (0.5,0.5,0.25)$, $\boldsymbol{r}_2 = (0.5,0.5,0.75)$ are the wave packet center positions and $\ell = 0.1$ is the width of packets. The electric field is set by $\be = \pm \hat{\boldsymbol{z}} \times \bb$ with opposite sign for each wave packet.
The amplitude $\xi$ characterizes the size of largest perturbation imposed on the background field,
\begin{eqnarray}
\xi \simeq \max \left|\frac{\bb - B_0\hat{\boldsymbol{z}}}{B_0}\right|.
\end{eqnarray}
\begin{figure}[t]
\includegraphics[width=0.5\textwidth]{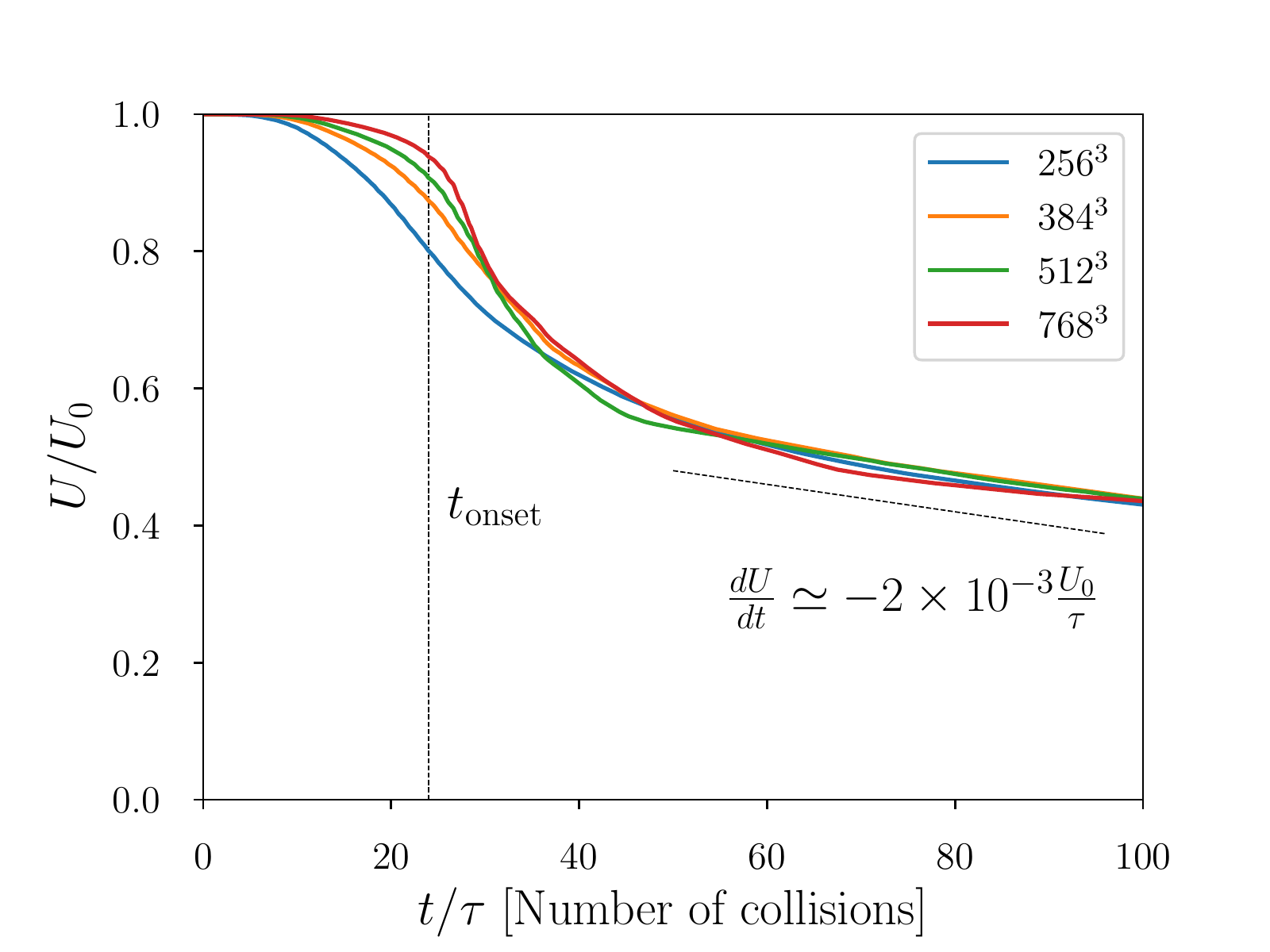}
\caption{Free energy evolution for different resolutions in 3D simulations of colliding \alfven wave packets with amplitude $\xi=0.5$.}
\label{energy3D}
\end{figure}

We observe that in 3D simulations, the collision of counter-propagating \alfven waves results in a forward energy cascade, in which energy is dissipated primarily by grid-heating (Channel (iv) in Section \ref{dissipation-channels}). This interpretation is supported by (1) consistency of the overall energy dissipation rate with increasing grid resolution, (2) observation of a definite time $t_{\rm onset}$ at which energy dissipation commences, and (3) formation of a Kolmogorov-type energy spectrum.

Figure \ref{energy3D} shows the time series of electromagnetic free energy $U(t)$ for the same model, $\xi = 0.5$, at different grid resolutions. All of the simulations exhibit an initial phase with slow dissipation lasting $t_{\rm onset}\sim 24 \tau$, a fast dissipation phase between $\sim 24\tau$ and $\sim 40\tau$,
and a subsequent gradual relaxation phase. The difference between the initial slow and fast dissipation phases becomes more pronounced as the grid resolution increases; the rate of energy dissipation prior to $t_{\rm onset}$ diminishes with increasing numerical resolution. Meanwhile, the energy lost by the system at late times $>40\tau$ is independent of the grid resolution to within roughly $5\%$.

\begin{figure}[t]
\includegraphics[width=0.5\textwidth]{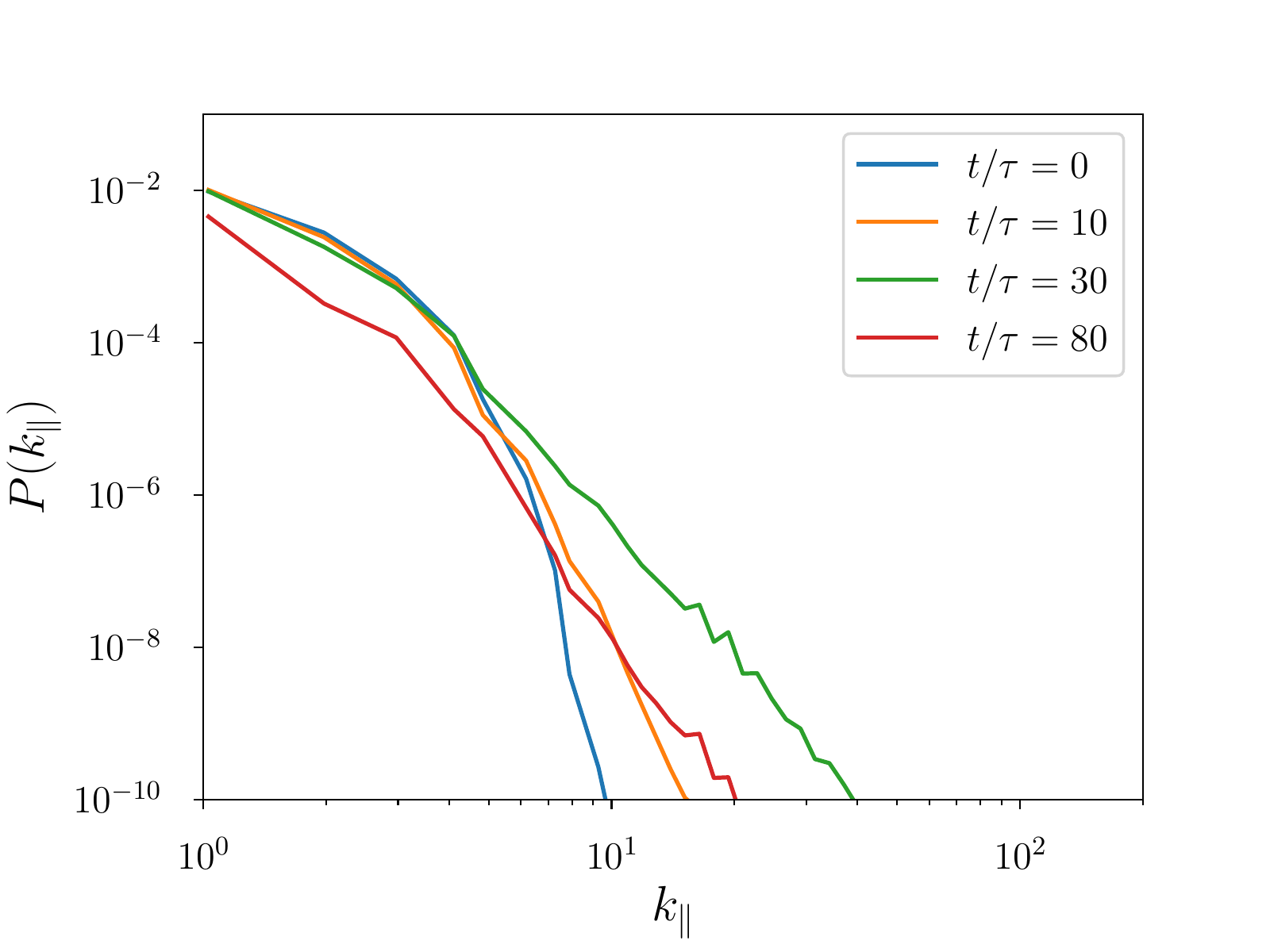}
\includegraphics[width=0.5\textwidth]{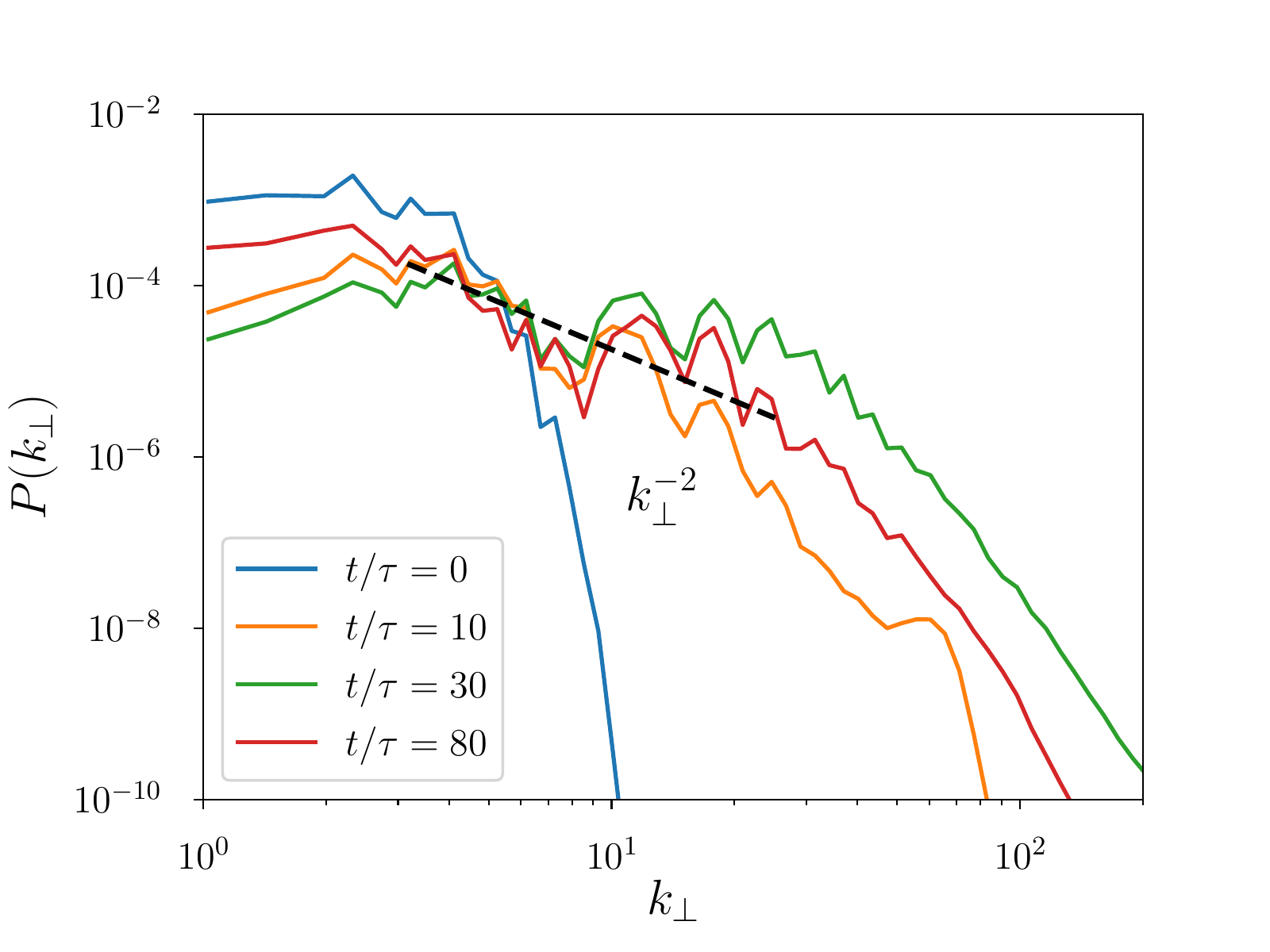}
\caption{Development of turbulent spectrum (snapshots at $t/\tau=0$, 10, 30, 80) in the simulation with the initial packet amplitude $\xi=0.5$ and grid resolution $512^3$. Upper panel: spectrum in $k_\parallel$ (parallel to the background field). Lower panel: spectrum in $k_\perp$(perpendicular to the background field).
The dashed line indicates the slope $P(k_\perp)\propto k_\perp^{-2}$.
}
\label{spec3Dev}
\end{figure}
Figure \ref{spec3Dev} shows the power spectrum evolution for a simulation with resolution $512^3$. Over time, the system develops waves at progressively increasing wavenumber, indicating a forward energy cascade. The spectrum extends to a maximum wavenumber $k_{\rm max}(t)$, which 
is seen to increase between $t=0$ and $t_{\rm onset}$.
At $t_{\rm onset}$, $k_{\rm max}$ reaches the nominal dissipation wavenumber $k_{\rm diss} \sim N / 10$, where $N$ is the number of grid points in each $(x,y,z)$ direction.
Figure \ref{spec3Dev} also reveals that the spectrum is significantly anisotropic, with $P(k_\perp) > P(k_\parallel)$ at all but the largest scales, indicating that energy cascades primarily in the direction perpendicular to the background field.
As energy moves from large to small (perpendicular) scales, the energy around large $k_\perp$ increases monotonically up until $t_{\rm onset}$, at which time modes around $k_{\rm diss}$ become significantly populated. Subsequently, some energy is reflected back toward low wave numbers, causing the power at large scales to grow between $t_{\rm onset}$ and $80\tau$.
The perpendicular spectrum eventually relaxes to a power-law consistent with $k_\perp^{-2}$ at $\sim 80\tau$. Such spectral slope is consistent with the so-called weak MHD wave turbulence spectrum, as reported by \cite{2001JETP...93.1052K}.

\begin{figure}[t]
\includegraphics[width=0.5\textwidth]{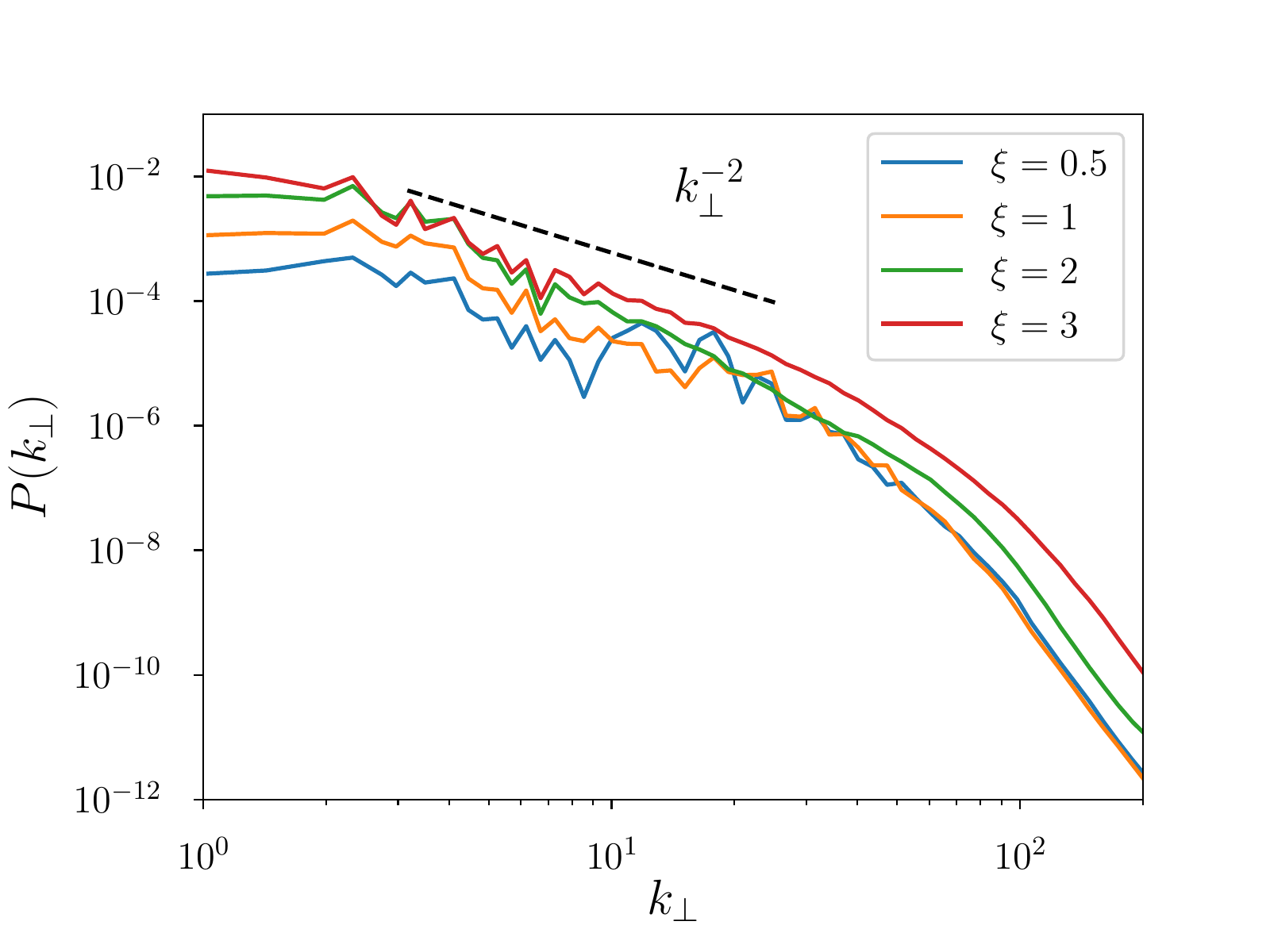}
\caption{
Turbulence spectrum in $k_\perp$ at $t=80\tau$ in four simulations with different initial packet amplitudes $\xi$.
}
\label{spec3Dst}
\end{figure}

The perpendicular power spectrum exhibits oscillations in $k_\perp$ at times earlier than $80\tau$.
This is due to a known feature of wave turbulence \citep{2011LNP...825.....N}, that energy is transferred mainly through resonant interactions; only a discrete set of secondary modes are excited by the primary modes. The resonant secondary modes then couple with the primaries and further drive the same secondaries. This leads to disproportionate energy transfer between particular sectors of the $k$-space, enhancing the energy concentration around preferred wavenumbers.
Figure \ref{spec3Dst} shows the perpendicular power spectrum after 80 collisions for various amplitudes $\xi$ of the initial wave packets in the range $0.5 - 3$.
The slopes of the perpendicular spectra are all close to $k_\perp^{-2}$.
We observe that the spectral oscillations are weaker for larger values of $\xi$.
This is because the strength of nonlinear interactions increases with $\xi$ and energy is distributed across a larger number of modes within a given time.

The existence of a universal time $t_{\rm onset}$ at which dissipation commences is consistent with a forward energy cascade, and a spectral energy distribution having finite \textit{energy capacity}, meaning that
\begin{equation}
	\lim \limits_{k_{\rm max} \rightarrow \infty} \int \limits_{k_0}^{k_{\rm max}} P(k) dk < \infty \, .
\end{equation}
This is the case for any power-law spectra $P(k)$ steeper than $k^{-1}$.
Such spectra have the property that the energy stored at wavenumbers higher than $k$ asymptotes toward zero with increasing $k$.
As energy cascades toward smaller scales, $k_{\rm max}$ must either increase without bound (thus exciting modes at the dissipation scale, however small), or some of the energy must be reflected toward larger scales.
We do see evidence in Figure \ref{spec3Dev} for such energy reflection, as the power around $k_\perp \sim 4$ first drops, but then rises again at $t \sim t_{\rm onset}$.
However, the energy distribution subsequently equilibrates to a Kolomogorov spectrum, with energy transferring continuously into the dissipation range, and leading to the divergence of $k_{\rm max}$.
The rapid increase of $k_{\rm max}$ implies that $t_{\rm onset}$ becomes insensitive to $k_{\rm diss}$, and thus to the grid resolution.
Therefore the energy spectrum $P(k_\perp) \propto k_\perp^{-2}$ seen in 3D simulations is compatible with the time series in Figure \ref{energy3D} which suggests a universal value of $t_{\rm onset}$.
In the next section we will show that 2D settings exemplify the opposite behavior, where the spectrum is very shallow, having infinite energy capacity.
Those 2D systems will not display numerical consistency of the dissipation onset time.

\subsection{2D simulations}
In our 2D simulations, the field is independent of the $y$ coordinate.
We utilize initial conditions
$\bb = B_0\hat{\boldsymbol{z}}+ B_y\hat{\boldsymbol{y}}$ in the $x-z$ plane, where two \alfven wave packets have Gaussian profiles,
\begin{equation}
\label{eqn:2d-wave-packets}
	B_y = \xi \sum\limits_{i=1,2} \exp\left(- \frac{|\boldsymbol{r}-\boldsymbol{r}_i|^2}{\ell^2} \right) \, .
\end{equation}
Here, $\boldsymbol{r}_1 = (0.5,0.25)$ are $\boldsymbol{r}_2 = (0.5,0.75)$ are the center positions of the two wave packets in the $x-z$ plane.
The width of the wave packet is the same ($\ell=0.1$) as in our 3D simulations.
The electric field is again $\be = \pm \hat{\boldsymbol{z}}\times \bb$, with opposite sign for each wave packet. The wave packets travel toward one another along the guide field (in the $\pm \hat{\boldsymbol{z}}$ directions), and have a cylindrical envelope in which the magnetic field is perturbed along the cylinder axis $\hat {\boldsymbol{y}}$. 
\begin{figure}[t]
\includegraphics[width=0.5\textwidth]{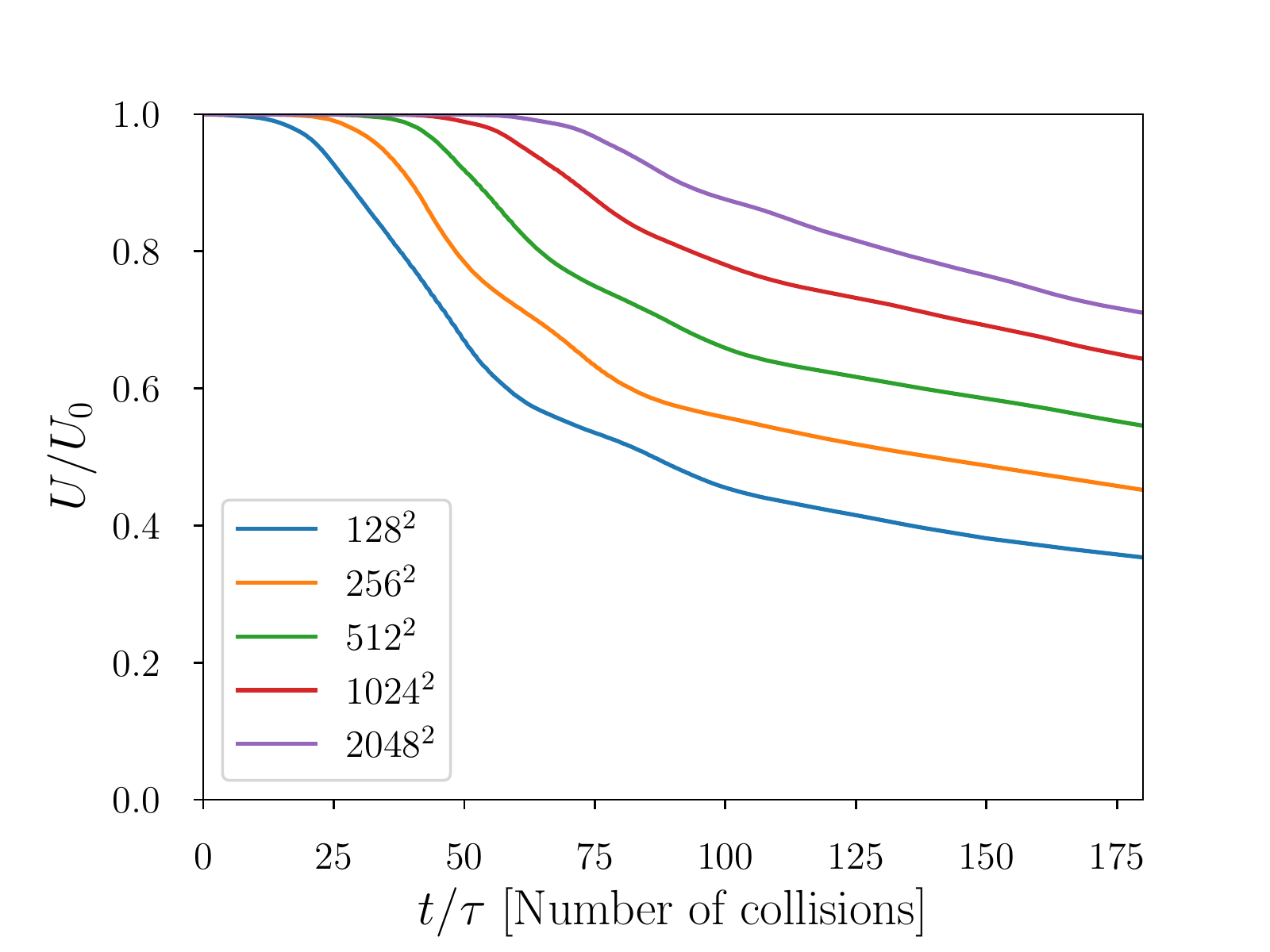}
\caption{ Free energy evolution for different resolutions in the 2D model with packet amplitude $\xi=0.4$.}
\label{energy2D}
\end{figure}

In 2D simulations, we observe that collisions between counter-propagating wave packets also result in a forward energy cascade.
However, unlike in the 3D case, 2D systems do not exhibit consistency of the overall dissipation rate for different grid resolutions. 
Figure \ref{energy2D} shows the time series of electromagnetic free energy for amplitude $\xi = 0.4$, with different numerical resolution.
The amount of energy $\Delta U$ dissipated before a given time $t_0$ is a decreasing function of the grid resolution, showing no trend toward a universal value.
Moreover, the onset of dissipation occurs at later times; there is no asymptotic $t_{\rm onset}$ that was observed in Section \ref{sec:3d} for the 3D case.
\begin{figure}[t]
\includegraphics[width=0.5\textwidth]{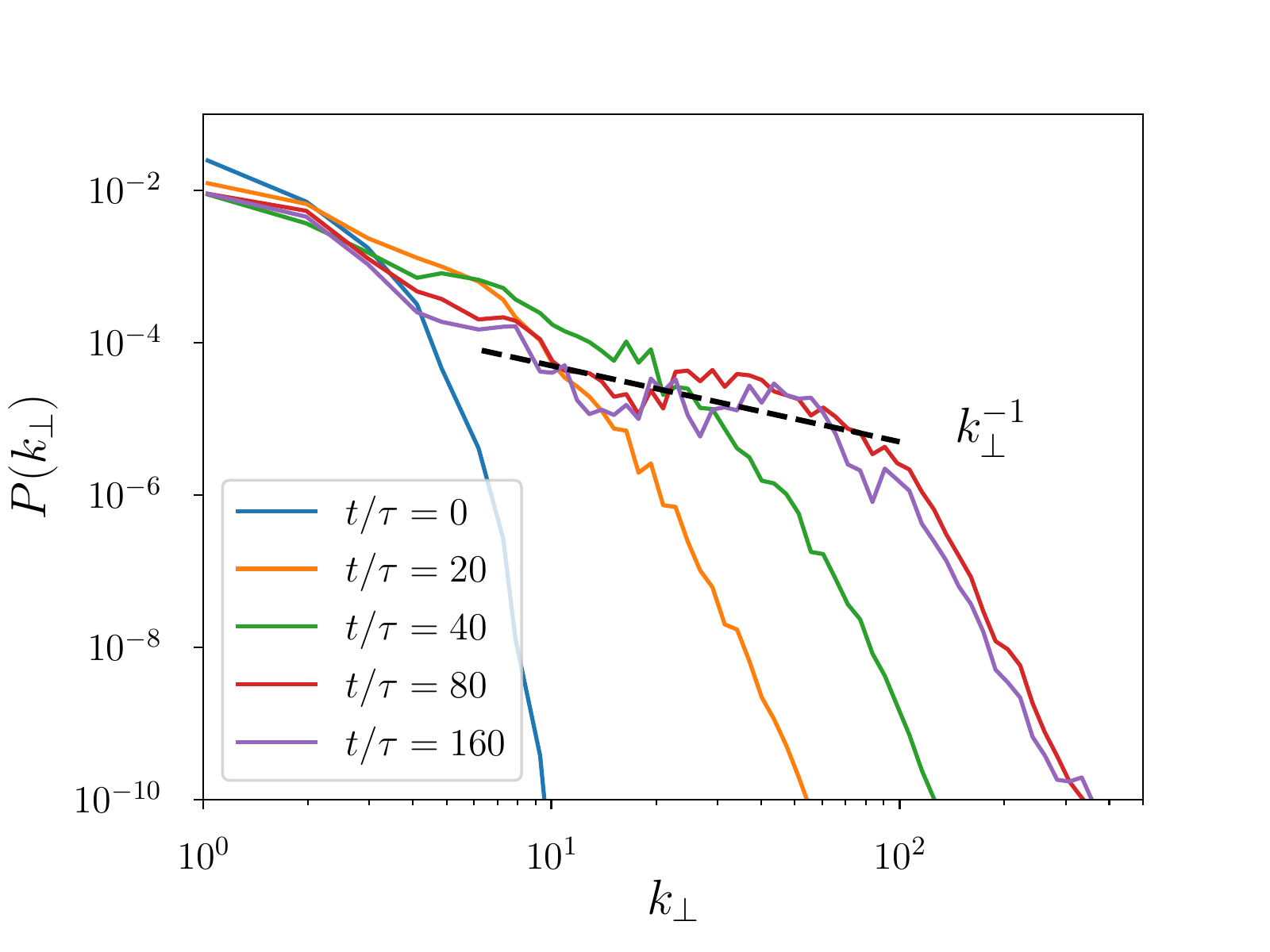}
\caption{Spectrum evolution for the 2D simulation with $2048^2$ resolution and amplitude $\xi =0.4$. The dashed line indicates the slope $P(k_\perp)\propto k_\perp^{-1}$.}
\label{spec2Dev}
\end{figure}

The energy spectrum in 2D simulations is also different from the 3D case. Figure \ref{spec2Dev} shows the spectral evolution for a model with $\xi = 0.4$ and grid resolution of $2048^2$. Over the course of tens of collisions, energy is gradually redistributed toward smaller scales, with a perpendicular spectrum $P(k_\perp) \propto k_\perp^{-1}$, significantly shallower than the 3D case.

The different energy dissipation rates seen in 2D versus 3D settings can be explained by the difference in their spectral slopes.
In particular, the energy spectrum in 2D has an unbounded energy capacity, because
\begin{equation}\label{eqn:energy-cap}
	U_0 = \alpha \int\limits_{k_{0}} ^{k_{\rm max}} \md k_{\perp} \; k_\perp^{-1} = \alpha 
    \,\ln
    \frac{k_{\rm max}}{k_0}
\end{equation}
would diverge if $k_{\rm max}\rightarrow \infty$. Here $\alpha$ is a normalization factor which may evolve with time. As energy cascades toward smaller scales, $k_{\rm max}$ increases, but remains finite. This fact is consistent with the increasing delay of the dissipation onset with increasing resolution, as it takes longer for $k_{\max}$ to reach $k_{\rm diss}$.

The evolution of the $k_\perp^{-1}$ turbulence spectrum is determined by the evolution of its normalization $\alpha(t)$.
Suppose that $k_{\rm max}$ increases as a power law with time, 
\begin{equation}
  k_{\rm max} \propto t^q, \qquad q>0.
\end{equation}
The normalization $\alpha$ must decrease as $k_{\rm max}$ increases,
\begin{equation}\label{eqn:alpha-of-t}
	\alpha(t) = \frac{U_0}{\ln(k_{\rm max}/k_0)} = \frac{U_0}{q\ln t+\beta},
\end{equation}
where $\beta$ is a constant. When $k_{\rm max}$ reaches $k_{\rm diss}$, grid heating begins to remove energy on scales smaller than $k_{\rm diss}^{-1}$, and the turbulence energy $U$ decreases below $U_0$,
\begin{equation}\label{eqn:U-of-t}
	U(t) = \alpha(t) \int\limits_{k_{0}} ^{k_{\rm diss}} \md k_{\perp}\; k_\perp^{-1} = \alpha(t) \ln \frac{k_{\rm diss}}{k_0}\, .
\end{equation} 
Equations \ref{eqn:alpha-of-t} and \ref{eqn:U-of-t} together yield the relation
\begin{equation}\label{eqn:U-model}
	\frac{U_0}{U(t)} = \frac{q\ln t + \beta}{\ln(k_{\rm diss} / k_0)}.
\end{equation}
This description assumes that $\alpha(t)$ (or the value of $q$) is independent of grid dissipation at high $k_\perp$. The value of $k_{\rm diss}$ is proportional to the grid resolution $N$ and the evolution of $U$ depends on $N$.

The predicted relation~(\ref{eqn:U-model}) can be tested by measuring $U(t)$ in the simulations with different resolutions and checking (1) whether $U_0/U(t)$ is indeed a linear function of $\ln t$, and (2) whether $q$ indeed has a universal value. This test is shown in Figure~6 for five different resolutions $N$ that span a factor of $16$. In each case, after $k_{\max}$ reaches $k_{\rm diss}$ we observe a linear growth of $U_0/U(t)$ with $\ln t$. We have measured its slope $s$ as a function of $\ln k_{\rm diss}$ and then calculated $q$ from the linear realtion inferred from Equation~(\ref{eqn:U-model}),
\begin{equation}
   \frac{1}{s}=\frac{1}{q}\ln\frac{k_{\rm diss}}{k_0}.
\end{equation}
For all five resolutions, the values of $(\ln k_{\rm diss},\,1/s)$ are found to follow the same line with $q\approx 1.75$, confirming the above analytical picture of the turbulence spectrum evolution. In contrast to the 3D simulations, dissipation slows down with increasing $N$.

\begin{figure}[t]
\includegraphics[width=0.5\textwidth]{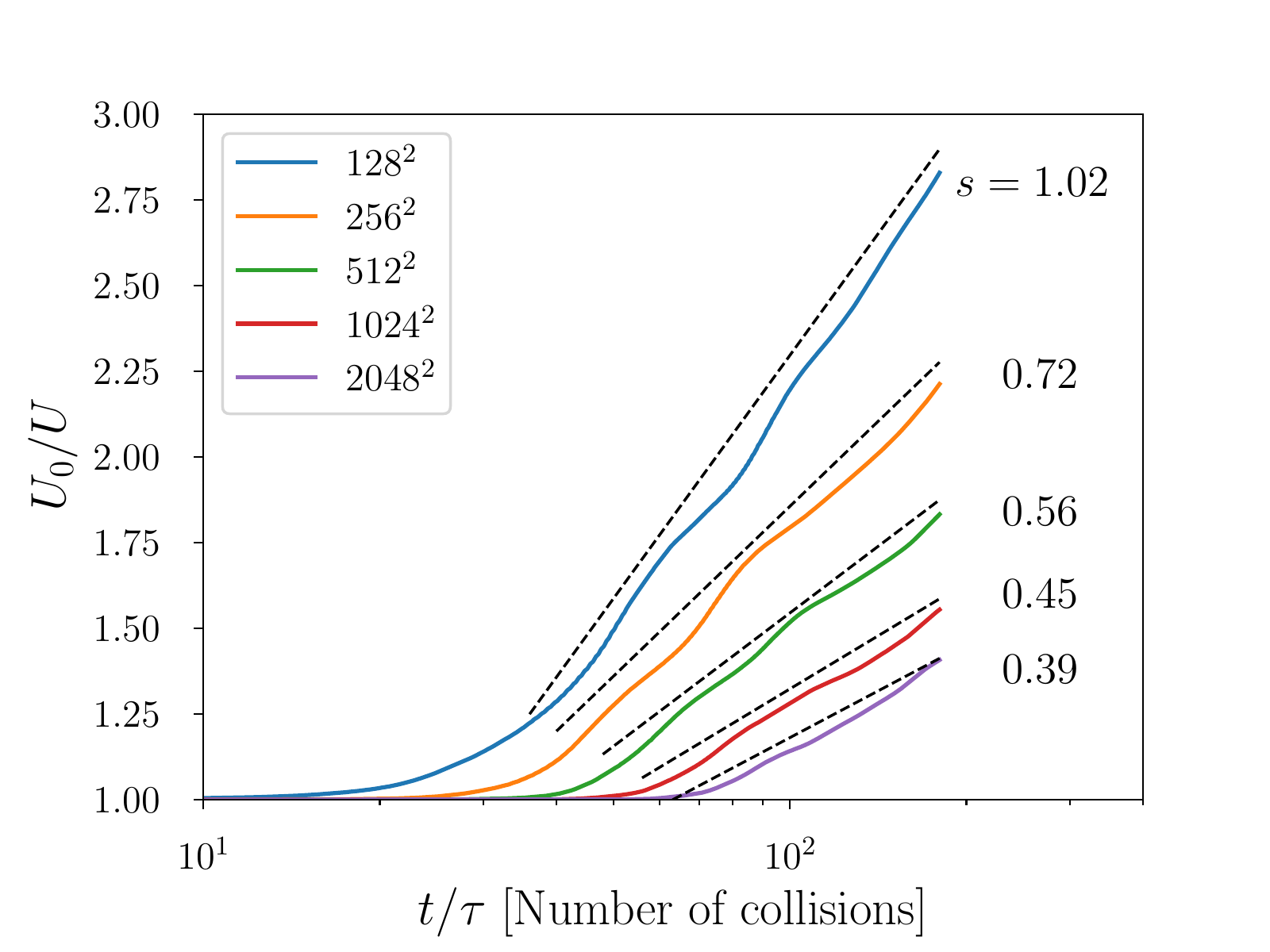}
\caption{ 
The evolution of $U_0/U(t)$ in 2D simulations. Dashed lines show the best-fit slopes of the linear relation between $U_0/U$ and $\ln t$. The slope value $s$ is indicated next to each curve.}
\label{fit}
\end{figure}

\section{Local dissipation and escape of waves}
\label{sec:fate}

As discussed in Section \ref{waves}, nonlinear interactions between \alfven waves can excite fast modes. These modes are not ducted along the magnetic field lines; they have group velocity in any direction and may escape the magnetosphere. In this section we examine the competition between the two energy sinks: local dissipation of the turbulent cascade and the escape of generated fast waves. We use the 3D setup of colliding \alfven wave packets as described in Section~3, and compare two sets of simulations, with and without wave escape, as explained below.

\subsection{Turbulent dissipation rate without wave escape}
\begin{figure*}
\includegraphics[width=0.5\textwidth]{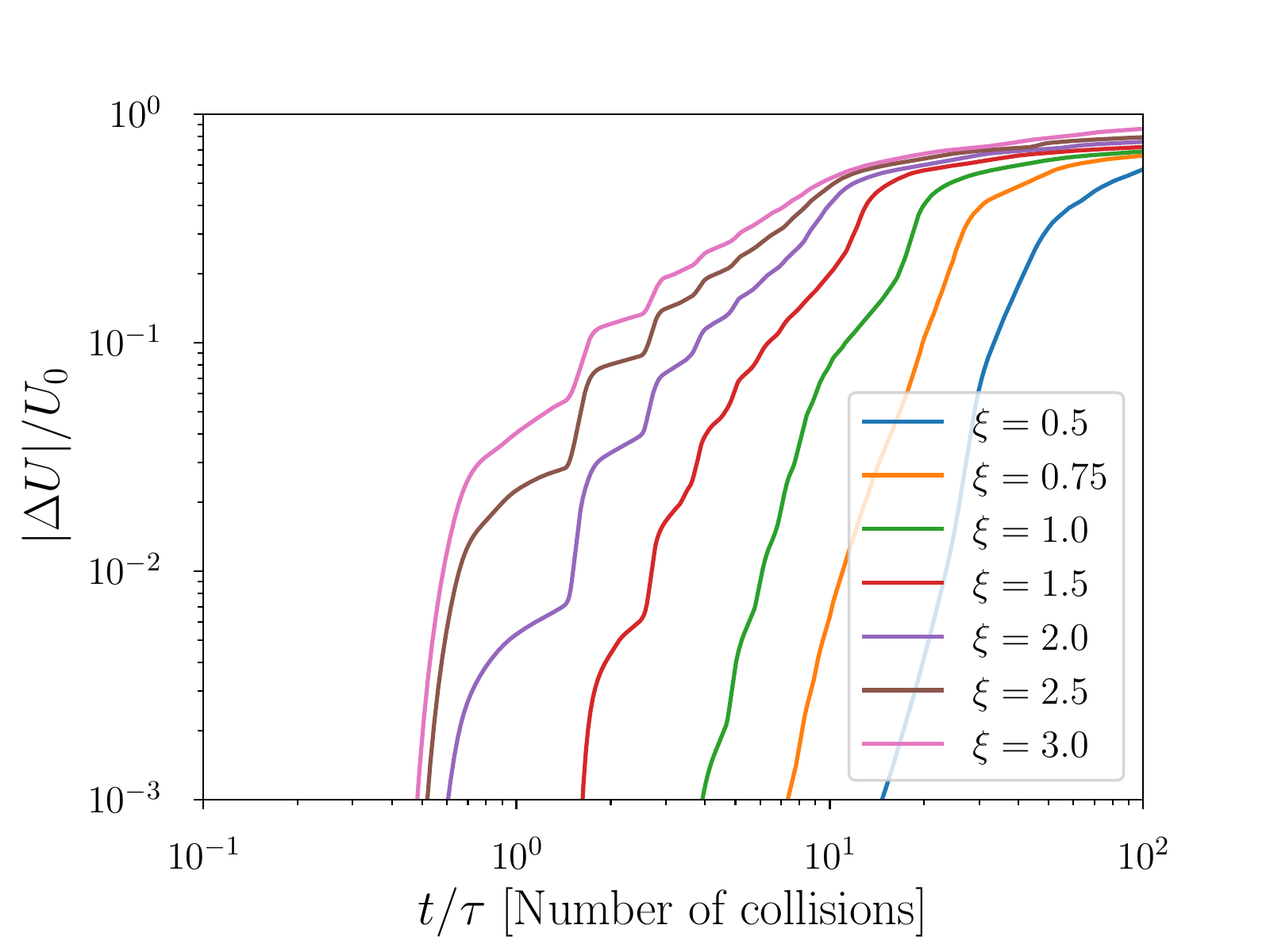}
\includegraphics[width=0.5\textwidth]{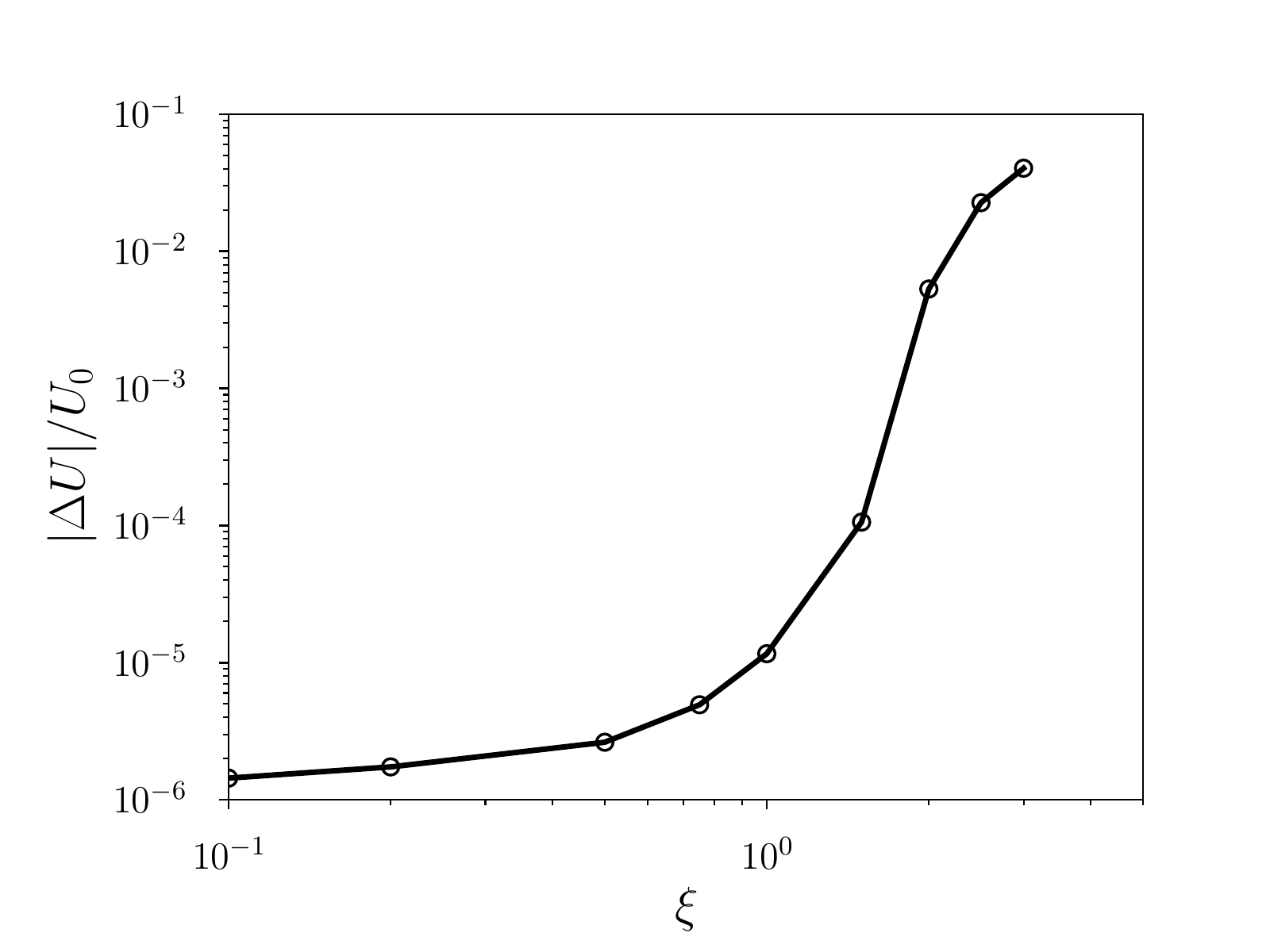}
\caption{Dissipation of \alfven wave packet energy $U_0$ in 3D simulations.
\textit{Left} -- Time series for the dissipated $f(t)=|\Delta U|/U_0$. Different curves show models with different packet amplitudes $\xi$.
\textit{Right} -- Dissipated energy fraction $|\Delta U|/U_0$ after the first collision as a function of packet amplitude $\xi$.
}
\label{casraterate3D}
\end{figure*}
The periodic boundary conditions for all ($x,y,z$) directions imply that waves cannot escape the computational box; they can only dissipate. The dissipation efficiency depends on the amplitude of the colliding \alfven packets $\xi$ and their sizes $\ell$. We have studied this dependence by calculating models with seven different values of $\xi$ between 0.5 and $3$, at fixed resolution of $512^3$. The results are presented in Figure~ 7, which shows evolution of the dissipated energy fraction,
\begin{equation}
  f(t)=\frac{U_0-U(t)}{U_0}.
\end{equation}
One can see that $f(t)$ is small in the first collisions, and its time dependence is step-like, because dissipation occurs only during the collisions,
when the two packets overlap. (The  duration of overlap is significantly shorter than the time between the collisions, because the packet width $\ell$ is smaller than the computational box size.) At later times, the field line bundle carrying the two packets becomes increasingly filled with strong \alfven turbulence, capable of dissipating energy outside the packets; then the dissipation curve $f(t)$ becomes smoother.

As expected, $f(t)$ is higher for the simulations with larger packet amplitudes $\xi$, because of the higher effectiveness of the nonlinear interaction.
The dissipated fraction after the first collision, $f_1=f(\tau)$, is plotted in the right panel of Figure \ref{casraterate3D} for different values of $\xi$. 
We find that $f_1$ is a very sensitive function of $\xi$, rising sharply from $10^{-5}$ at $\xi = 1$ to $10^{-2}$ at $\xi = 2$. 

\subsection{Turbulent cascade with escaping fast modes}
To evaluate the energy radiated away by fast modes, we have introduced a ``sponge'' layer along the transverse domain boundaries, intended to absorb fast waves reaching the boundaries.
The elimination of energy transported by fast waves into the sponge layer simulates their escape.

The sponge layer is implemented by adding an Ohmic-like dissipation term $-\sigma_s \be$ to the force-free current in Equation \ref{current}. This term leads to exponential damping of the electric field on the timescale $\sigma_s^{-1}$. 
We adopt a spatial profile of $\sigma_s(x,y)$ that leads to faster damping of the electric field near the boundary of the computational domain in $x$-$y$ plane ($0<x<1$; $0<y<1$),
\begin{equation}
	\sigma_s = \frac{1}{2 \tau} \left( 1 - e^{-8 \delta^4} \right), \qquad  \delta = \max \left(\frac{r_\perp - r_0}{d - r_0}, 0 \right),
\end{equation}
where $r_\perp = \sqrt{(x - d)^2+(y - d)^2}$, $d = 0.5$ is half of the transverse domain scale, and $r_0 = 0.3$ is the distance from the $z$-axis within which absorption is switched off completely, $\sigma_s=0$.
Near the boundary, the damping time scale $\sigma_s^{-1}$ drops to $2 \tau$ (equal to the light crossing time of the computational box).
The energy dissipated in the sponge layer is a proxy for energy escaping the system in the form of fast waves.

\begin{figure}[t]
\includegraphics[width=0.5\textwidth]{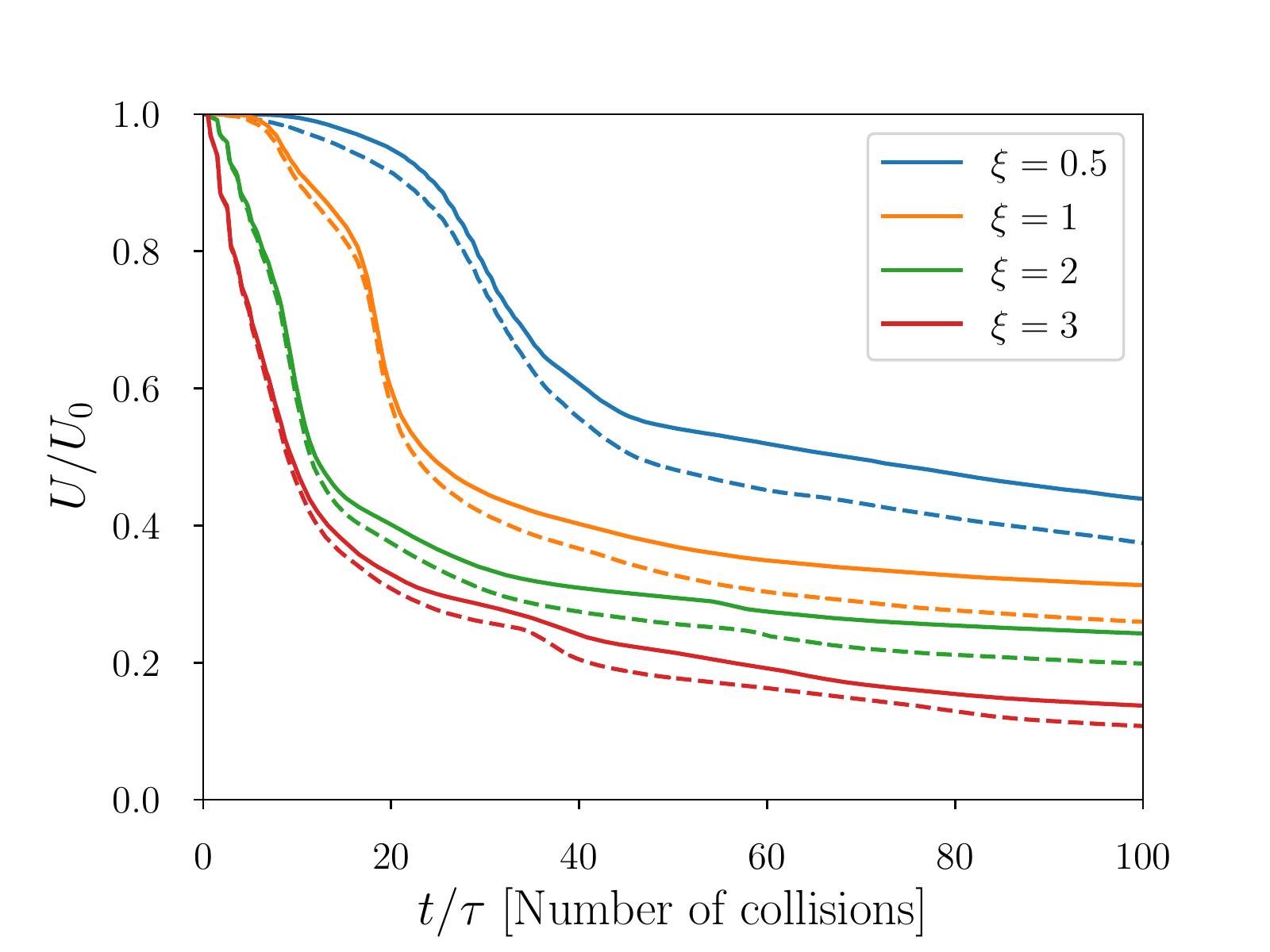}
\caption{Comparison between free energy evolution $U(t)$ in the simulations with (dashed curves) and without (solid curves) damping of fast modes at the boundary. 
The difference between solid and dashed curves shows the effect of fast mode escape compared with local dissipation through the turbulent cascade.
}
\label{3Dfastwave}
\end{figure}

The loss of fast modes at the boundaries leads to a faster decline of the free energy in the box $U(t)$ compared with the simulations without wave escape.
The magnitude of this effect is a measure of the effectiveness of the energy loss through the boundaries compared with local dissipation through the turbulent cascade.
Figure \ref{3Dfastwave} shows the comparison of four pairs of simulations with and without the sponge layer (identical otherwise). All eight simulations have resolution $512^3$. The four different amplitudes $\xi=0.5, 1, 2, 3$ are chosen to investigate how the competition between wave escape and turbulent damping depends on the initial amplitudes of the packets.

For instance, in the simulation with $\xi = 0.5$, during the first $10$ collisions (prior to $t_{\rm onset}$), fast waves carry away $\sim 2\%$ of the initially available free energy $U_0$, compared to $\sim 1\%$ taken by turbulent dissipation.
Radiation of fast waves is thus the primary energy loss channel prior to the onset of developed turbulence. After turbulence is fully developed, at times $\gtrsim t/\tau=100$, 
the energy radiated by fast waves accounts for only $6\%$ of $U_0$ while turbulent dissipation accounts for nearly $60\%$.

A smaller energy fraction is carried away by fast waves in the simulations with larger $\xi$.
In the run with $\xi=3$, the sponge layer accounts for only $3\%$ of $U_0$.
This trend is the result of a stronger coupling of fast waves in nonlinear interactions.
As the fast waves participate in the turbulence to a greater degree, they are scattered and damped more efficiently, dissipating more energy locally in the magnetosphere.

\section{
Enhanced immediate dissipation and FFE failure
}
\label{sec:breakffe}

The results of the preceding sections show that wave damping takes many crossing times $\tau$, even at very high amplitudes $\xi>1$. Since this conclusion is based on numerical simulations with a concrete setup, one would like to know whether the conclusion is robust. To address this question we have tried to vary the initial setup of the \alfven wave packets in search for more efficient damping, and found that in some cases our FFE simulations predict much quicker dissipation, which occurs immediately in the first collisions of the wave packets, even before the development of turbulence.

\subsection{Immediate dissipation observed in FFE simulations}
The immediate dissipation effect is sensitive to the initial polarization of the wave packets. It is maximized (and most convenient to study) in the simplest 1D ``slab'' setup. Then the initial perturbations of the magnetic field in the two colliding packets, $\bb_1$ and $\bb_2$, can point in any two chosen directions in the $x$-$y$ plane perpendicular to the guide field $\bb_0$. The angle between $\bb_1$ and $\bb_2$ will be denoted by $\theta$. The relative polarization $\theta$ is an important parameter of the packet collision problem, in addition to the packet amplitude $\xi$.

The simple 1D setup is better suited for the study of polarization effects than the 3D and 2D setups of the preceding sections. Recall that in the 3D setting the spherical packets could not have a single direction for $\bb_1$ or $\bb_2$, and thus the polarization angle was not well defined. In the 2D setting explored in Section~4.2, we chose $\bb_1$ and $\bb_2$ along the $y$-axis perpendicular to the simulation plane $x$-$z$, which allowed us to confine the packets in a circular region in the $x$-$z$ plane. However, the requirement of $\bb_{1,2}$ being parallel (or anti-parallel) to the $y$-axis leaves only two possibilities for the relative polarization, $\theta=0$ or $180^\circ$, and in Section~4.2, we stuck to the case of $\theta = 0^\circ$.
Therefore, in both 3D and 2D simulations presented in Section~4 we observed dissipation only through turbulence cascade to the grid scale, which takes a significant time.

The reason for immediate damping discussed in the present section is the activation of the dissipation channel (iii) listed in Section~3.5. Our FFE simulations show, for some values of $\theta$ and $\xi$, field evolution that violates the condition $E<B$, and then the procedure of enforcing this condition (Section~3.4) creates strong dissipation.

One can see the role of relative polarization $\theta$ for this effect from a simplified consideration that neglects the nonlinear character of  packet collisions and merely looks at the linear superposition of the colliding packets. When $\bb_1$ and $\bb_2$ are parallel ($\theta=0$), the magnetic fields of the packets add constructively while their electric fields add destructively --- the opposite Poynting fluxes of the two packets $\be_1\times \bb_1$ and $\be_2\times\bb_2$ require them to have antiparallel electric fields $\be_1$ and $\be_2$. 
By contrast, when the packets have nearly anti-aligned $\bb_1$ and $\bb_2$, the electric fields $\be_1$ and $\be_2$ become parallel and add constructively making it possible for $E$ to exceed $B$ for sufficiently large amplitudes $\xi$.

A simple estimate gives the range of $\theta$ and $\xi$ where this effect may be expected.  Let us consider two counter-propagating packets of amplitude $\xi$ centered at $z_1(t)$ and $z_2(t)$. The packets can have, for example, a Gaussian shape, $B_{1,2}=\xi B_0\exp[-(z-z_{1,2})^2/\ell^2]$. Let us choose the $y$-axis along $\bb_1$; then
\begin{eqnarray}
   \bb_1&=&\xi \exp\left[-\frac{(z-z_1)^2}{\ell^2}\right]\,\boldsymbol{b}_1, \quad 
       \boldsymbol{b}_1=(0,1,0) \\
   \bb_2&=&\xi \exp\left[-\frac{(z-z_2)^2}{\ell^2}\right]\,\boldsymbol{b}_2, \quad 
       \boldsymbol{b}_2=(\sin\theta,\cos\theta,0).
\end{eqnarray}
where we use the units $B_0=1$, . The corresponding electric fields are $\be_{1,2}=\mp \hat{\boldsymbol{z}} \times \bb$,
so the angle between the electric fields is $180^\circ - \theta$.
If the non-linear interaction of the packets is neglected, then at the point of maximum overlap (at $z=z_1=z_2$) the superposed field magnitudes would be
\begin{eqnarray}
	(\bb_1+\bb_2)^2 &=& 2\xi^2(1+\cos\theta) + 1 \\ \nonumber
	(\be_1+\be_2)^2 &=& 2\xi^2(1-\cos\theta) \, .
\end{eqnarray}
Magnetic dominance would thus be lost when
\begin{equation}\label{cr1}
	-4\xi^2\cos\theta >1 \, .
\end{equation}
%
This condition can be satisfied if $\theta>90^\circ$, and is easiest to satisfy if $\theta=180^{\circ}$. In the latter case, a moderately strong amplitude $\xi>0.5$ is required.
\begin{figure}[t]
\includegraphics[width=0.5\textwidth]{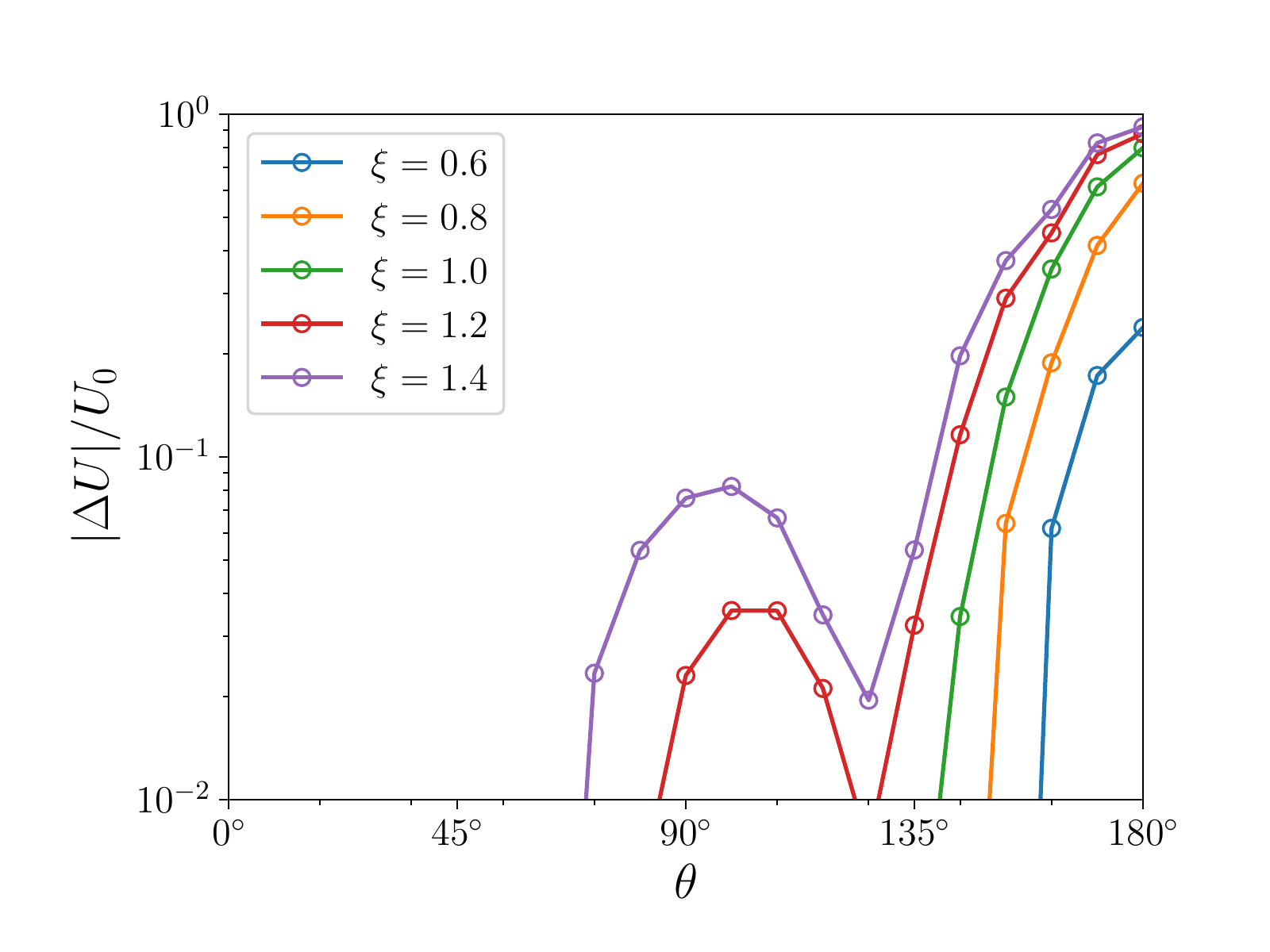}
\caption{Dissipated energy fraction in the 1D wave packet collision as a function of the relative polarization angle $\theta$, for various amplitudes $\xi$ of the colliding packets.}
\label{dis1d}
\end{figure}
%
\begin{figure}[htpb]
\includegraphics[width=0.5\textwidth]{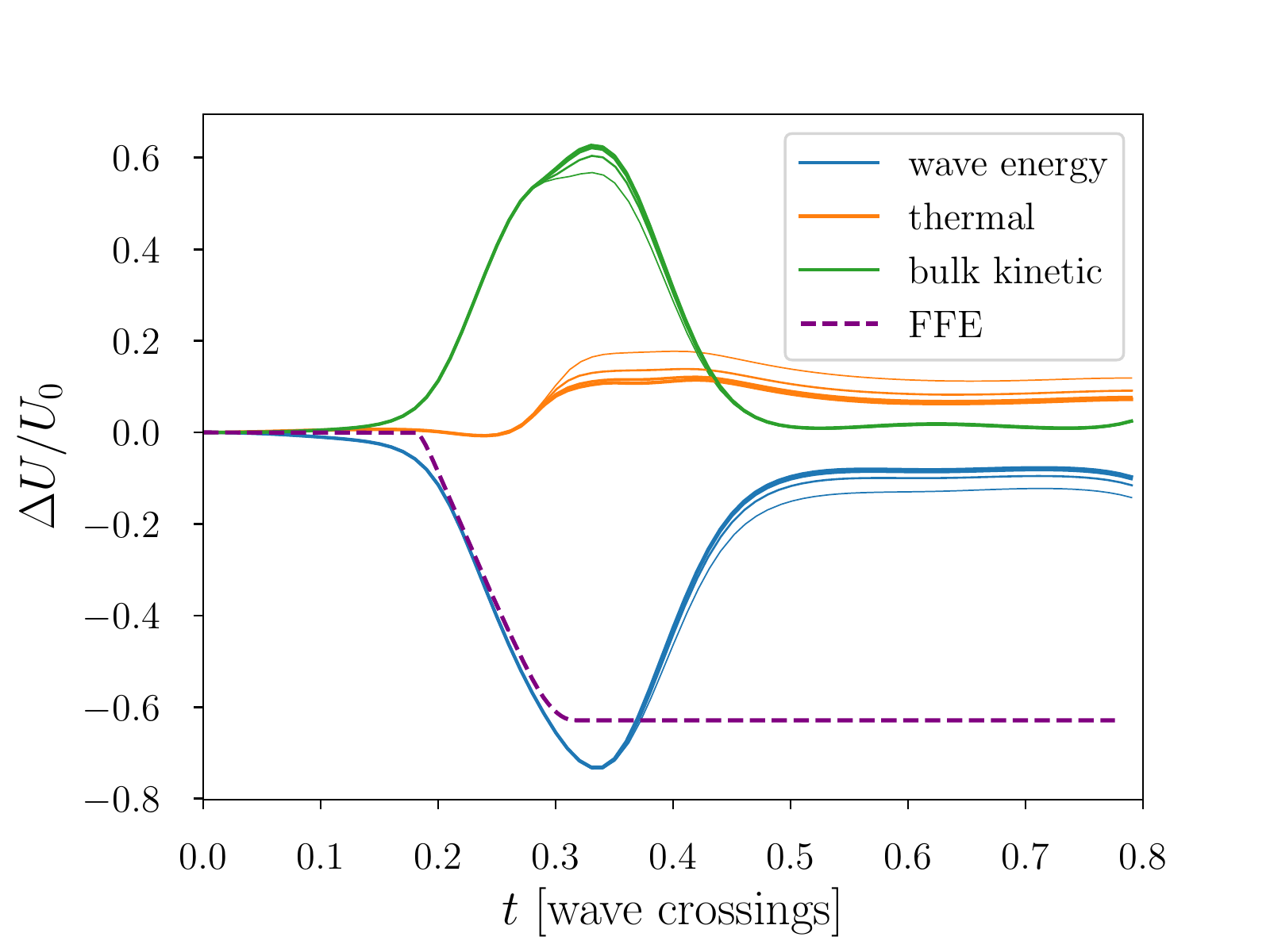}
\caption{
Comparison of the FFE simulation (dashed curve) with the full MHD simulation that tracks both plasma energy and electromagnetic energy (solid curves). The colliding \alfven wave packets have amplitude $\xi=0.8$ and relative polarization angle $\theta = 180^\circ$. Numerical convergence of the MHD simulation is shown by plotting the results obtained with resolutions $N=256,512,1024,2048$ (the increasing line thickness corresponds to the increasing resolution).
}
\label{rmhd}
\end{figure}

In general, such large amplitude waves interact non-linearly and their magnitude cannot be determined by linear superposition. Remarkably however, we observe from Equation \ref{current} that in FFE the non-linear terms vanish for 1D anti-polarized plane waves counter-propagating along $\bb$. 
Thus, when $\theta = 180^\circ$, the wave packets pass through one another unchanged, as long as $E<B$, and the linear superposition estimate for the loss of $E<B$ condition should be accurate.

\begin{figure*}[htpb]
\includegraphics[width=\textwidth]{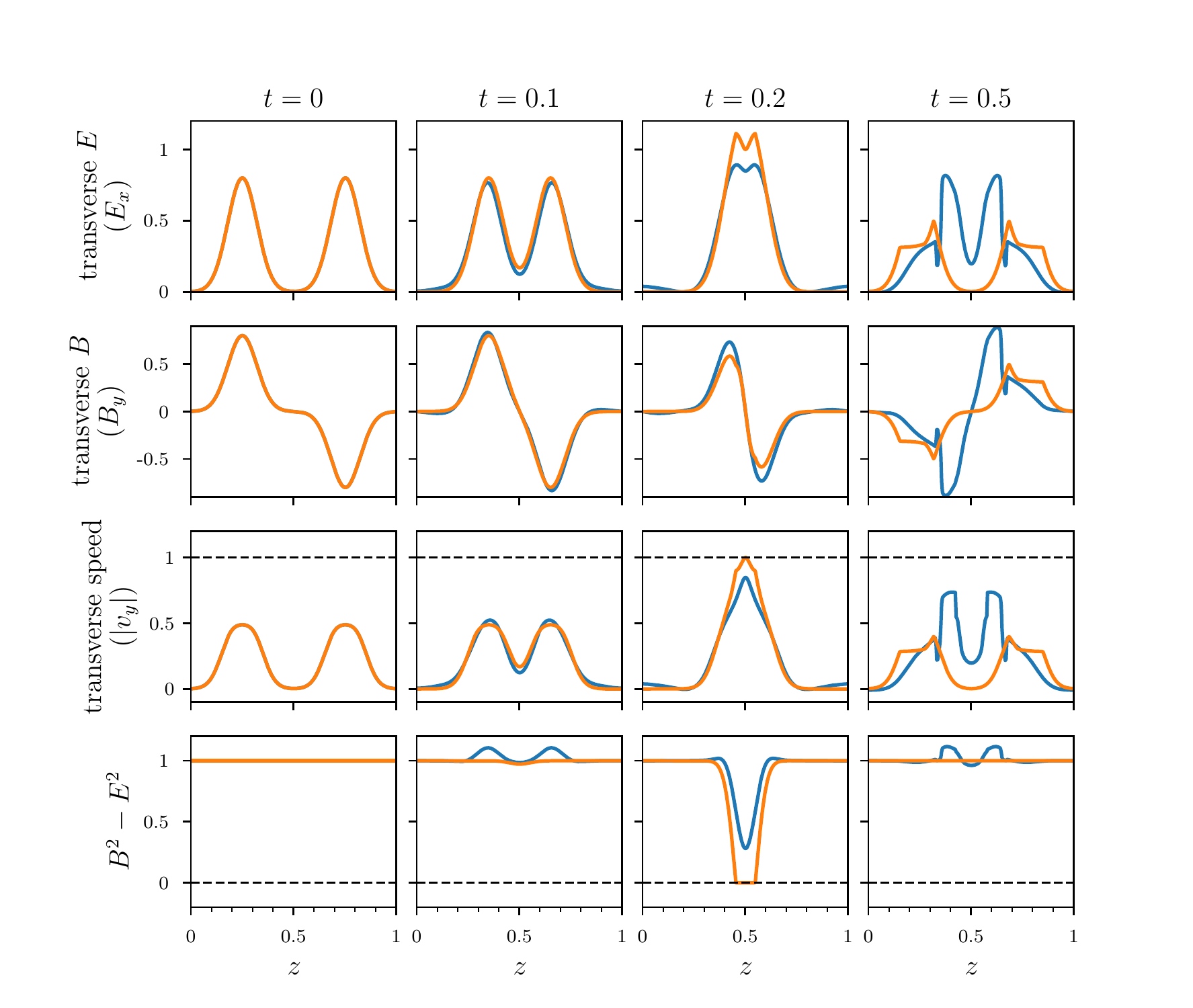}
\caption{
Snapshots of the FFE (orange) and MHD (blue) simulations shown in Figure~10. 
The snapshots are taken at four times, from $t=0$ (left) to $t=0.5$ (right) in units of the light crossing time of the computational box. 
}
\label{snapshots}
\end{figure*}

These expectations are tested in Figure~9, which presents the simulation results for the 1D setup described above. It shows the fraction of energy dissipated after a single collision of the two packets, $f_1=|\Delta U|/U_0$, for different values of the relative polarization angle $\theta$ and packet amplitude $\xi$. We find that strong dissipation, caused by the imposed shortening of the electric field to sustain $E<B$, becomes active for nearly anti-polarized packets if $\xi \gtrsim 0.5$. For example, we see 20\% energy loss after a single collision when $\xi = 0.6$ and $\theta = 180^\circ$. As $\xi$ increases to 1.4, we observe $f_1$ growing to $\sim 90$\% for $\theta$ approaching $180^\circ$, and exceeding 10\% for $\theta>135^\circ$.
 
Figure \ref{dis1d} also reveals a second peak in $f_1(\theta)$ near $\theta=90^\circ$. This peak is not predicted by the above linear superposition estimate, and thus has a nonlinear origin. However, it is also caused by the violation of $E<B$ condition.
It can be understood by looking at the evolution of $\be\cdot \bb$ in the linear superposition approximation, $\be\cdot \bb = 2\xi^2\sin\theta$. 
The linearly superposed fields would violate the FFE condition $\be\cdot \bb=0$, and nonlinear effects are responsible for sustaining this condition: the system is forced to generate a longitudinal electric field $E_z \propto \xi^2$, and for large enough $\xi$ this electric field component leads to the loss of magnetic dominance. 
This effect is proportional to $\sin\theta$ and thus strongest at $\theta\approx 90^\circ$.

\subsection{MHD simulations and the spurious character of immediate dissipation in FFE}
One may conclude from the FFE simulations that the collisions of large-amplitude \alfven waves with favorable polarization 
gives strong immediate dissipation. The dissipation mechanism in this case is the result of the customary procedure of shortening $E$ to sustain $E<B$. 
However, we point out that there is no guarantee that this procedure correctly captures the true field evolution. 
The true behavior of the system when $E$ reaches $B$ is outside the realm of FFE and can be understood only with a more complete physical model. The model must explicitly include a component of the system that takes the energy (and momentum) lost by the field.

Therefore, we have performed similar simulations of the 1D packet collisions in the full relativistic MHD, which does not neglect the plasma stress-energy tensor, and conserves the total energy and momentum of field and plasma. Plasma moves with a subliminal velocity $\boldsymbol{v}$, and MHD simulations never break the condition $E<B$, since the electric field $\be = -\bV \times \bb$ is obtained from the primitive variables, rather than evolved independently as it is in FFE. We use the relativistic MHD code \texttt{Mara} \citep{2012ApJ...744...32Z}.

MHD is expected to approach the FFE regime in the limit of high magnetization $\sigma\gg 1$. Therefore, in our simulations we choose a high $\sigma=25$, where $\sigma=B_0^2/\rho_0$ is defined for the background magnetic field $B_0$ and $\rho_0$ is the initial (uniform) plasma density in the computational box. 
Otherwise, the simulation setup is the same as described in Section~6.1 for the 1D FFE simulations.

Figure \ref{rmhd} compares the MHD and FFE results. For this test we chose the case where the colliding packets have amplitude $\xi=0.8$ and are anti-polarized ($\theta=180^\circ$).
We see that MHD and FFE predict similar evolution of the free electromagnetic energy $U(t)$ up through the peak of the collision. 
Then the MHD evolution strongly deviates from the prediction of the FFE simulation.
Importantly, the electromagnetic energy lost in the MHD simulation is compensated by a gain in the plasma kinetic energy, while the FFE code removes that same energy $\Delta U$ from the simulation irreversibly by the $E < B$ fix. 

Note that the electromagnetic fields of the two packets initially carry a significant $y$-momentum of the same sign. The MHD simulation shows that during the collision a large fraction of this momentum is taken by the plasma. The plasma momentum density is enhanced at the interface between the colliding packets by a factor $\sim 20$, comparable to $\sigma$. This enhancement is caused by two factors: the plasma is compressed by a factor of $\sim 6$ and the Lorentz factor of its transverse drift ($\boldsymbol{v}_\perp=\be\times\bb_0\parallel -\hat{\boldsymbol{y}}$) exceeds 3.
As a result, a large fraction of the packet electromagnetic energy temporarily converts into bulk kinetic energy of the plasma accelerated along the $y$-axis.
Once the collision is over, the accelerated plasma is stopped by magnetic stresses, restoring the electromagnetic field energy nearly to its initial value. 

Figure~\ref{snapshots} shows in more detail the evolution of the field and plasma in the FFE and MHD simulations.
Unlike the FFE, the MHD system does not reach the ``floor'' $B^2-E^2=0$, because it would correspond to the drift speed equal to the speed of light and hence infinite kinetic energy of the plasma. The plasma is strongly accelerated when $B^2-E^2$ is reduced, and the subsequent dynamics are completely different in the two simulations.
We conclude that the strong dissipation effect observed in FFE simulations is spurious. It is caused by the failure of FFE simulations to keep track of energy that is temporarily removed from the electromagnetic field when $E$ approaches $B$.





\section{Discussion}\label{sec:discussion}

The fate of \alfven waves excited in a magnetar magnetosphere is interesting from observational point of view if the wave energy eventually converts to radiation.  In particular, the hot plasma fireball formed in giant flares could be powered by dissipation of waves \citep{1995MNRAS.275..255T}, or the waves may be absorbed by the neutron star and feed surface afterglow emission \citep{2015ApJ...815...25L}. In the relativistic magnetospheres, where the magnetic field dominates over the plasma rest mass and all waves propagate with the speed of light, the nonlinear behavior of wave turbulence is poorly known. In this paper we employed numerical simulations to systematically study turbulence excited by colliding packets of \alfven waves and the resulting dissipation. The packets are assumed to be launched by an unspecified triggering event (e.g. a fast displacement of the crust or a global magnetospheric instability), which determines the initial packet amplitude $\xi = \delta B / B_0$ and size $\ell$.

\subsection{Summary of results}

Most of our results are obtained from high-resolution FFE simulations in a Cartesian box, using the 5th order conservative finite differencing scheme described in \cite{2011MNRAS.411.2461Y}. 
We have also explored situations where the FFE approximation becomes insufficient; then we employed relativistic MHD simulations. Our results are as follows.

(1) Our 3D simulations of packet collisions show that significant dissipation begins when a turbulence cascade develops down to the grid scale. 
The cascade is dominated by modes with wavevectors orthogonal to the background magnetic field, $k_\perp\gg k_\parallel$, and its
spectrum is steep, with a slope close to $-2$. We observed consistency of the cascade and the resulting dissipation rate with increasing grid resolution, and concluded that dissipation of the 3D turbulence is well modeled by ``grid heating'' (the removal of high-$k$ modes on the grid scale). 
The simulations reveal that even for wave packets of enormous amplitudes $\xi=1-3$ dissipation develops slowly, over many (10-100) collisions of the packets bouncing in the magnetosphere. The main reason for the dissipation delay is the relatively slow development of the broad spectrum of high-frequency modes and the onset of a persistent energy flow in the cascade down to the dissipation scale.

(2) We have found that it is essential to calculate the wave turbulence in three dimensions. Similar simulations restricted to two dimensions (where the fields are assumed to be independent of one  coordinate running transverse to the guide field $\bb_0$) are deficient. They produce qualitatively different results and show no convergence with increasing resolution, because the 2D cascade has a flat spectrum with an infinite energy capacity.

(3) \alfven waves are trapped, because they are ducted along the magnetic field lines, however their collisions generate fast modes that can carry energy away across the field lines. We have measured energy loss due to fast mode escape and found this effect to be weaker than energy dissipation on the field lines carrying the \alfven waves.

(4) When two strong \alfven waves collide, the electromagnetic field can experience immediate significant energy loss. This effect is qualitatively different from the cascade dissipation on the grid scale. It occurs when the Lorentz invariant $B^2-E^2$ is pushed to zero during the field evolution in some parts of the colliding packets, threatening to violate the condition $E<B$. This effect is possible only for certain relative polarizations of the colliding waves and is maximum when the waves have anti-aligned magnetic fields, as demonstrated by a simple 1D model. Our simulations revealed that $B^2-E^2$ can be pushed to zero also in a collision of waves with orthogonal polarizations; this occurs due to a non-linear effect responsible for sustaining $\be\cdot\bb=0$.

(5) We have shown that the permanent energy loss in FFE simulations caused by $B^2-E^2\rightarrow 0$ is spurious.
FFE has no component of the system other than the electromagnetic field and thus has no choice but to permanently remove the energy lost by the field. Our relativistic MHD simulations revealed that in fact this energy is temporarily stored in the
plasma that is compressed and
accelerated to a high Lorentz factor perpendicular to the background magnetic field $\bb_0$.
As the two colliding wave packets finish their interaction, the relativistic plasma motion is eventually decelerated and most of its energy is returned to the electromagnetic field. 

Our results indicate that damping of \alfven waves in the magnetosphere is surprisingly slow even at extremely high amplitudes, and so the waves can bounce in the magnetosphere for many crossing times. We have discussed the physical reasons for this behavior, and conclude that the slow damping is likely a true feature, not an artifact of our approximations. However, one should bear in mind the following simplifications adopted in our simulations.  
\\
(1) Our computational box was rectangular and filled with a uniform background magnetic field $\bb_0$. Magnetic field lines in a real magnetosphere are curved and can extend far from the star, where the field is much weaker. \alfven waves bouncing on such extended field lines will significantly increase their amplitudes as they propagate  in the outer weak-field region.
\\ 
(2) We focused on magnetospheres with energy density $B^2/8\pi$ much greater than the plasma rest mass. This regime almost always holds for the magnetosphere of 
a neutron star. However, during a giant flare, a significant fraction of the magnetic energy may be dissipated and stored in the electron-positron fireball trapped in the magnetosphere. Then the plasma inertia can become a significant factor in the evolution of \alfven wave turbulence.
\\
(3) Our simulations assumed perfect reflection at the boundaries that represent the stellar surface in the computational box. Since the two colliding packets are symmetric in our simulation setup, their perfect reflection at the surface is equivalent to periodic boundary conditions. In reality, the reflection coefficient is slightly below unity, and $\sim 10$\% of the packet energy is transmitted into the star \citep{2015ApJ...815...25L}. 
\\
(4) We described the spurious immediate dissipation in packet collisions using only 1D (FFE and MHD) simulations. As one can see from Figure~\ref{snapshots}, a short-lived current sheet forms at the packet collision interface, in the $x$-$y$ plane perpendicular to $\bb_0$. If the current sheet becomes tearing unstable, magnetic reconnection will occur and  dissipate some energy. The tearing is not allowed in 1D models, and so 2D or 3D simulations are required to investigate the possible reconnection in the current sheet. As a first step, we have ran several test 2D simulations using kinetic code TRISTAN-MP. 
We found that magnetic reconnection is important only when the packet amplitude $\xi$ is much larger than unity. We leave the detailed study of magnetic reconnection 
in packet collisions to a future paper.

\subsection{Fate of wave energy in magnetar flares}

One implication of our results is that dissipation of \alfven waves in the magnetosphere is less efficient than their absorption by the neutron star. \citet{2015ApJ...815...25L} showed that $\sim 10$ interactions of the wave packet with the stellar crust is sufficient to absorb a large fraction of its energy. That work simulated \alfven wave packets hitting the neutron star crust with realistic density profile $\rho(z)$ and obtained the reflection and transmission coefficients for this interaction, $\mathcal{R}$ and $\mathcal{T}$. The numerical results were also found consistent with an analytical estimate for wave tunneling into the crust using WKB approximation. For typical magnetar fields $B_0>10^{14}$~G  and sizes of the wave packet $\ell\sim 10$~km (comparable to the star radius), the transmission coefficient is $\mathcal{T}=10-20\%$. It increases for stronger $B_0$, because it implies a higher \alfven speed inside the magnetar crust.

The shear wave transmitted into the heavy crust is much slower than the magnetospheric \alfven wave. It continues to propagate into the deeper and denser crustal layers with a decreasing speed and a diminishing amplitude. 
However the strain in the wave {\it grows} as $\propto \rho^{1/4}$ and eventually induces a plastic flow. As a result, the wave energy converts to heat, melting the solid material at the bottom of the liquid ocean, which is $\sim 100$~m deep in magnetars. Thus, most of the magnetospheric wave energy is expected to convert to heat at $\sim 100$~m below the stellar surface. \citet{2015ApJ...815...25L} also calculated how the heat diffuses from this depth and is mostly lost to neutrino emission; a fraction $\sim 0.1$ of the heat will reach the surface and feed the surface afterglow weeks to months after the event that triggered the magnetospheric waves.

Only a fraction $\fdiss$ of  the magnetospheric wave energy will be dissipated locally in the magnetosphere (and an even smaller fraction $\fesc$ will convert to fast modes that escape the field lines carrying the \alfven waves). In particular, in our simulations, the wave energy fraction dissipated per collision of packets is $ \simlt 1\%$, which is $\simgt 10$ times lower than $\mathcal{T}$. Therefore, we roughly estimate $\fdiss\simlt 0.1$. It may still be interesting for powering fireball radiation. However, a more promising source for fireball energy appears to be magnetic reconnection in a global instability of the over-twisted magnetosphere, as observed in the simulations of Parfrey et al. (2013). The reconnection event immediately dissipates significant energy. It also launches strong waves, which dissipate with efficiency $\fdiss$ in the magnetosphere, but mostly disappear into the star and feed its invisible neutrino emission and a delayed afterglow from the stellar surface.

\acknowledgements
X.L. and J.Z. appreciate useful input from Maxim Lyutikov, in particular the suggestion to simplify the description of FFE eigen modes by using the temporal gauge.
A.M.B. is supported by NASA grant NNX17AK37G and a grant from the Simons Foundation \#446228.

\appendix
\section{Resonant Three Wave Interactions}

The second order nonlinear current $\bj^{(2)}_{\rm nl}$ in Equation \ref{Maxwell2} reads
\begin{eqnarray}\label{J2}
	\bj^{(2)}_{\rm nl} &=& \nabla\cdot\be^{(1)}\frac{\be^{(1)}\times\hat{\bz}}{B_0} + \frac{\hat{\bz}\cdot\nabla\times\bb^{(1)}}{B_0}\bb^{(1)} \nonumber\\
	&&+\frac{\bb^{(1)}\cdot\nabla\times\bb^{(1)} - \be^{(1)}\cdot\nabla\times\be^{(1)}}{B_0}\hat{\bz}-2\hat{\bz}\cdot(\nabla\times\bb^{(1)})\frac{\hat{\bz}\cdot\bb^{(1)}}{B_0}\hat{\bz} \, .
\end{eqnarray}
Note that the terms in the second line of the above equation (those proportional to $\hat{\bz}$) only source $A_z$,and thus do not excite any propagating waves. 

Let us consider the interaction of a pair of linear waves 
$\bA^{(1)}_{1}$ and $\bA^{(1)}_{2}$. 
Then we substitute into Equation~(\ref{J2}) $\be^{(1)}$ and $\bb^{(1)}$ obtained from $\bA^{(1)}=\bA^{(1)}_{1}+\bA^{(1)}_{2}$. This yields the second order current,
\begin{equation}\label{J2b}
	\bj^{(2)}_{\rm nl} = \nabla\cdot\be_1^{(1)}\frac{\be_2^{(1)}\times\hat{\bz}}{B_0} + \frac{\hat{\bz}\cdot\nabla\times\bb_1^{(1)}}{B_0}\bb_2^{(1)} + (1\leftrightarrow 2) + (\mathrm{terms \ proportional \ to \ } \hat{\bz}) \, ,
\end{equation}
where $(1\leftrightarrow 2)$ means the repetition of previous terms but with subscript $1$ and $2$ exchanged.
We seek a solution $\bA^{(2)}$ of Equation~(\ref{Maxwell2}) which is sourced by $\bj_{\rm nl}^{(2)}$, is itself an eigen mode, and whose amplitude grows in time. Our ansatz is thus $\bA^{(2)}(\boldsymbol{r}, t) = \Lambda_m(t)\boldsymbol{e}_m\exp[i(\boldsymbol{k}^{(2)}\cdot\boldsymbol{r}-\omega^{(2)}t)]$ where $\omega^{(2)}$ and $\boldsymbol{k}^{(2)}$ satisfy either the fast or \alfven wave dispersion relations.
Inserting $\bA^{(2)}(\boldsymbol{r}, t)$ into Equation~(\ref{Maxwell2}), we obtain the evolution equation for the wave amplitude $\Lambda_m(t)$,
\begin{equation}\label{ampevol}
	\partial^2_t \Lambda_m(t) - 2i\omega^{(2)} \partial_t \Lambda_m(t) =\omega^{(2)} \bj^{(2)}_{\rm nl}\cdot\boldsymbol{e}_m e^{i(\omega^{(2)}t-\boldsymbol{k}^{(2)}\cdot\boldsymbol{r})} \, .
\end{equation}
Inspection of Equation~(\ref{J2b}) reveals that $\bj^{(2)}_{\rm nl}$ is proportional to $\exp[i(\boldsymbol{k}_{12} \cdot\boldsymbol{r}-\omega_{12} t)]$ where $\boldsymbol{k}_{12}=\boldsymbol{k}_1^{(1)}+\boldsymbol{k}_2^{(1)}$ and $\omega_{12}=\omega_1^{(1)}+\omega_2^{(1)}$. The right hand side of Equation~(\ref{ampevol}) may thus be written as
\begin{equation}\label{eqn:lambda-ode1}
	\partial^2_t \Lambda_m(t) - 2i\omega^{(2)} \partial_t \Lambda_m(t) =C_{12m} e^{i (\boldsymbol{k}_{12}-\boldsymbol{k}^{(2)})\cdot\boldsymbol{r}} e^{-i (\omega_{12}-\omega^{(2)})t}\, ,
\end{equation}
where $C_{12m}$ has no space or time dependence (these coefficients describe the strength of wave-wave interactions and are evaluated below for each of the allowed channels).
The absence of spatial dependence on the left hand side of 
Equation~(\ref{eqn:lambda-ode1}) implies that its right hand side is independent of $\boldsymbol{r}$, which requires $\boldsymbol{k}^{(2)}=\boldsymbol{k}_{12}$.
The temporal evolution of $\Lambda_m(t)$ satisfies the equation,
\begin{equation}\label{eqn:lambda-ode2}
	\partial^2_t \Lambda_m(t) - 2i\omega^{(2)} \partial_t \Lambda_m(t) = C_{12m} e^{-i (\omega_{12}-\omega^{(2)})t} \, .
\end{equation}
The general solution of Equation~(\ref{eqn:lambda-ode2}) subject to the initial condition $\Lambda_m(0)=0$ (and neglecting the constant of integration) is given by
\begin{equation}
	\Lambda_m(t) = C_{12m} \frac{1 - e^{-i(\omega_{12} - \omega^{(2)})t}}{\omega_{12}^2 - (\omega^{(2)})^2} \, .
\end{equation}
For arbitrary values of $\omega^{(2)}$, the amplitude
oscillates in time. However, as $\omega^{(2)} \rightarrow \omega_{12}$, the oscillation period grows longer, and when the resonance condition is met precisely, $\Lambda_m(t) \rightarrow i C_{12m} t / 2 \omega_{12}$. Energy transfer from the primary waves is only possible for such resonant interactions.

Below we list the expressions for $C_{12m}$ for each allowed resonant channel.
\begin{enumerate}
	\item $\mathcal{A}+\mathcal{A}\rightarrow\mathcal{F}'$
	\begin{equation}
C_{\mathcal{A}\mathcal{A}\mathcal{F}'}=\frac{-i \omega}{B_0\sqrt{\omega \omega_1\omega_2}} \frac{\Lambda_1\Lambda_2}{k_{\perp}k_{1\perp}k_{2\perp}}\left[\left( \omega_1\omega_2 - k_{1z}k_{2z} \right) \left(2k_{1\perp}^2k_{2\perp}^2 + (k_{1\perp}^2+k_{2\perp}^2)\boldsymbol{k}_{1\perp}\cdot \boldsymbol{
	k}_{2\perp} \right)\right]\, .
	\end{equation}
Here two interacting \alfven waves with frequencies $\omega_1(\boldsymbol{k}_1)$ and $\omega_2(\boldsymbol{k}_2)$, and amplitudes $\Lambda_1$ and $\Lambda_2$, generate a fast mode $\omega(k)$ that satisfies the resonance conditions $\boldsymbol{k}=\boldsymbol{k}_1+\boldsymbol{k}_2$ and $\omega=\omega_1+\omega_2$.

	\item $\mathcal{A}+\mathcal{F}\rightarrow\mathcal{F}'$
	\begin{equation}
    C_{\mathcal{A}\mathcal{F}\mathcal{F}'}=\frac{ i \omega}{B_0\sqrt{\omega \omega_\mathcal{F}\omega_\mathcal{A}}} \frac{\Lambda_{\mathcal{A}}\Lambda_{\mathcal{F}}}{k_{\perp}k_{\mathcal{A}\perp}k_{\mathcal{F}\perp}}\left[\left( \omega_\mathcal{A}\omega_\mathcal{F} - k_{\mathcal{A}z}k_{\mathcal{F}z} \right) k_{\mathcal{A}\perp}^2\left( \boldsymbol{k}_{\mathcal{F}\perp}\times \boldsymbol{k}_{\mathcal{A}\perp} \right)\cdot\hat{\bz} \right]\, .
	\end{equation}
Here an \alfven wave with frequency $\omega_{\mathcal{A}}(\boldsymbol{k}_{\mathcal{A}})$ and amplitude $\Lambda_{\mathcal{A}}$ interacts with a fast mode with frequency $\omega_{\mathcal{F}}(\boldsymbol{k}_{\mathcal{F}})$ and amplitude $\Lambda_{\mathcal{F}}$. The interaction generates a new fast mode $\omega(\boldsymbol{k})$ that satisfies $\omega=\omega_{\mathcal{A}}+\omega_{\mathcal{F}}$ and $\boldsymbol{k}=\boldsymbol{k}_{\mathcal{A}}+\boldsymbol{k}_{\mathcal{F}}$.
	 
    \item $\mathcal{A}+\mathcal{F}\rightarrow\mathcal{A}'$
	\begin{equation}
    C_{\mathcal{A}\mathcal{F}\mathcal{A}'}=\frac{ i \omega}{B_0\sqrt{\omega \omega_\mathcal{F}\omega_\mathcal{A}}} \frac{\Lambda_{\mathcal{A}}\Lambda_{\mathcal{F}}}{k_{\perp}k_{\mathcal{A}\perp}k_{\mathcal{F}\perp}}\left[\left( \omega_\mathcal{A}\omega_\mathcal{F} - k_{\mathcal{A}z}k_{\mathcal{F}z} \right) k_{\mathcal{A}\perp}^2\left( \boldsymbol{k}_{\mathcal{F}\perp}\cdot \boldsymbol{k}_{\mathcal{A}\perp} + k^2_{\mathcal{F}\perp} \right) \right]\, .
	\end{equation}
Here the interaction is similar to the previous one, except that the third (generated) wave $\omega(\boldsymbol{k})$ is an \alfven wave rather than a fast mode.
\end{enumerate}

\acknowledgments

\bibliography{ms}

\clearpage

\end{document}